\documentclass[%
 reprint,
 superscriptaddress,
 amsmath,amssymb,
 aps,
 pra,
floatfix
]{revtex4-1}

\usepackage{booktabs}
\usepackage{subcaption}
\usepackage{graphicx,epsf}
\usepackage{amsfonts}
\usepackage{amssymb}
\usepackage{amsmath}
\usepackage{overpic}
\usepackage{braket}
\usepackage{xcolor}
\usepackage{cancel}
\usepackage{fancyhdr}
\usepackage[normalem]{ulem}

\newcommand{\be}{\begin{equation}}
\newcommand{\ee}{\end{equation}}
\newcommand{\bea}{\begin{eqnarray}}
\newcommand{\eea}{\end{eqnarray}}

\def\lb{\label}

\newcount\bozza \bozza=1
\ifnum\bozza=1
\newdimen\shift \shift=-2truecm
\def\lb#1{%
{\label{#1}\rlap{\kern\shift{$\scriptstyle#1$}}}}
\else\def\lb#1{\label{#1}} \fi

\allowdisplaybreaks

\begin{document}

\title{Effects of intercomponent couplings on the appearance of the time-reversal symmetry breaking fermion-quadrupling state in  two-component London models}

\author{I. Maccari}
\affiliation{Department of  Physics, Stockholm University, Stockholm SE-10691, Sweden}
\author{E. Babaev}
\affiliation{Department of   Physics, The Royal Institute of Technology, Stockholm SE-10691, Sweden}

\begin{abstract}

Detection of a bosonic metallic state that breaks the $Z_2$ time-reversal symmetry has been recently reported in Ba$_{1-x}$K$_x$Fe$_2$As$_2$ with a doping level $x \approx 0.8$.
This is a metallic state of fermionic quadruplets that breaks time-reversal symmetry. As such, it has no condensed Cooper pairs but has a long-range order between fermionic quartets.
In the present manuscript, we investigate the emergence of this phase in a two-component London model via Monte Carlo simulations as a function of various intercomponent couplings. 

\end{abstract}
\maketitle

\section{Introduction}

Superconductivity arises as a consequence of the formation and condensation of electron  pairs. The condensation of Cooper pairs in many materials is well described by the mean-field Bardeen-Cooper-Schrieffer (BCS) theory ~\cite{Bardeen1957_microscopic, Bardeen1957_theory}.

However, multicomponent systems may exhibit a different kind of ordering associated with the long-range order of electron quadruplets~\cite{babaev2002phase,babaev2004superconductor,Smorgrav2005_observation,Bojesen2013time,Bojesen2014_phase,agterberg2008dislocations,berg2009charge,Herland2010,
cho2020z, fischer2016fluctuation,shaffer2021theory,Chung2022berezinskii}.
Related phases have also been discussed in strongly-correlated ultracold atoms \cite{kuklov2003counterflow,Kuklov2004,Dahl2008_preemptive} and other models \cite{Kuklov2006,Kuklov2008_deconfined}.
This type of order arises when the energy cost of composite topological defects (i.e. bound states of defects in different fields) is significantly cheaper than the energy cost associated with elementary topological defects. In this case, thermal or quantum fluctuations can
induce the proliferation of composite defects that break the phase coherence of Cooper pairs, without breaking the phase coherence of fermionic quartets. Thus leading to the appearance of a quartic fermionic state.

The recent experiment \cite{Grinenko2021_state} reported the observation of a fermionic-quadrupling condensate that spontaneously breaks time-reversal symmetry in Ba$_{1-x}$K$_x$Fe$_2$As$_2$.

The multiple broken symmetries required for the formation of this type of fermionic-quadrupling order can result from the presence of multiple bands crossing the Fermi surface that, in turn, can lead to a ground state where the inter-band phase difference is neither $0$ nor $\pi$~\cite{Ng2009, Stanev2010, Carlstrom2011_lengthscales, Maiti2013, silaev2017phase, Boeker2017, Grinenko2017, 
Kivelson2020, Grinenko2020,Grinenko2021_unsplit}. Such a ground state,
has an additional $Z_2$ two-fold degeneracy corresponding to the spontaneous breaking of the time-reversal symmetry (BTRS).
In this scenario, the superconducting (SC) ground state spontaneously breaks a $U(1) \times Z_2$ total symmetry.

The muon-spin rotation experiments~\cite{Grinenko2017, Grinenko2018} have found evidence for such a multicomponent $s + is$ superconducting state at low temperatures in Ba$_{1-x}$K$_x$Fe$_2$As$_2$ with a doping level $x \approx 0.8$.
The main evidence in favor of this state comes from the analysis of polarization of spontaneous magnetic fields \cite{Grinenko2018,Vadimov2018,garaud2014domain}, however the closely related $s+id$ state \cite{lee2009pairing} cannot be completely  ruled out today.

At the level of mean-field BCS theory, the critical temperature associated with the spontaneous breakdown of the time-reversal
symmetry is always smaller than or equal to the superconducting temperature: $T_c \geq T_c^{Z_2}$. \cite{Stanev2010,Carlstrom2011_lengthscales,Maiti2013,Boeker2017}.
The initial experimental results shown in~\cite{Grinenko2017, Grinenko2018} for a  certain range of doping, were consistent with this picture.
However, very recent experimental results ~\cite{Grinenko2021_state} found in Ba$_{1-x}$K$_x$Fe$_2$As$_2$ a regime  near the doping level $x \approx 0.8$ where $T_c^{Z_2}>T_c$,  signaling the onset of a fermionic-quadrupling phase  with spontaneously broken time-reversal symmetry. The conclusion was reached based on a number of experimental probes: muon-spin rotation experiments, conductivity, diamagnetic response, thermolelectric and ultrasound probes.
These probes revealed a very interesting state of matter that required further exploration. For a  recent theoretical work on the effective model, and some of the properties of this state see \cite{garaud2021skyrmions}.

The previous theoretical studies beyond mean-field approximation \cite{Bojesen2013time,Bojesen2014_phase,Grinenko2021_state}
indicate that the formation and size of that state depend on the strength of direct intercomponent coupling, the magnetic field penetration length relative to other length scales, and on the strength of mixed-gradient terms. However, systematic investigation of the interplay of these parameters was not performed. Here we present such an investigation.

The paper is organized as follows. In Sec. II, we introduce the continuum two-component London model. In Sec. III we discuss the Villain approximation and the details of the Monte Carlo (MC) simulations. In Sec. IV, we present the numerical results obtained in the different regimes investigated.  Conclusions are given in Sec. V. 
\section{The model}

As discussed in detail in~\cite{Garaud2017}, the free-energy density of a clean three-band superconductor with a bilinear intra-band Josephson interaction, can, under certain conditions, be approximated by a model with only two components and a biquadratic Josephson coupling.
Similar model arises for $s+is$ state 
generated by impurities in dirty two-band superconductors \cite{garaud2018properties}.
Finally, except for certain differences in structure of gradient terms, similar models also result from other systems such as $s+id$  \cite{lee2009pairing}, $p+ip$ \cite{Agterberg1998vortex,fischer2016fluctuation,Haugen2021first}, $s+ig$ \cite{Kivelson2020}
and systems that break time-reversal symmetry.

In this work, we will study  a  two-component London model in three spatial dimensions, as a model for $U(1) \times Z_2$ superconductors in the type-II limit.

The free-energy density of the system reads:
\begin{equation}
\begin{split}
    &f= \frac{1}{2} \left({\mathbf{\nabla}} \times \mathbf{A} \right)^2 + \sum_{i=1,2}\frac{\rho_i}{2} \left( {\mathbf{\nabla}} \phi_i - e \mathbf{A} \right)^2 + \\  -\nu& \left({\mathbf{\nabla}} \phi_1 -e \mathbf{A} \right) \cdot \left({\mathbf{\nabla}}\phi_2 -e\mathbf{A} \right)+ \eta_2 \cos[2(\phi_1  -\phi_2  )].
    \end{split}
    \label{continuum_London}
\end{equation}
Here $\phi_i \in [0,2\pi)$ are phase, coupled by the gauge field $\mathbf{A}$. The two condensates have the same electric charge: $e_1=e_2=e $.
The model \eqref{continuum_London} accounts for three different intercomponent interactions: the coupling of the two charged condensates via the fluctuating gauge field $\mathbf{A}$, the interaction via the coupling constant $\nu$, and the second-order biquadratic Josephson interaction with coupling constant $\eta_2$. 
The coefficient $\nu$
sets the intercomponent current-current coupling.  Such dissipationless-drag interaction  (i.e. mixed-gradient) terms are generically present in multicomponent systems and originate for example from Fermi-liquid corrections or strong correlations as shown in various physical contexts~\cite{leggett1975theoretical,
Sjoberg:76, Kuklov2004, Kuklov2006, Svistunov2015, Sellin.Babaev:18,hartman2018superfluid,linder2009calculation}.
 From Eq.~\eqref{continuum_London2}, one can easily derive the stability condition of the system, i.e. the condition ensuring the free energy to be bounded from below, being:  $\nu < \sqrt{\rho_1 \rho_2}$. 
 
 Without loss of generality, in what follows we will fix $\rho_1=1$, tuning the disparity of the components via the parameter $\alpha=\rho_2/\rho_1$. This corresponds to a reduction of the free parameters of the model by a proper rescaling of the coupling constants, the gauge field, the electric charge and the free energy. 

In the absence of the Josephson interaction, the model \eqref{continuum_London} has a $U(1) \times U(1)$ symmetry.
Phase transitions in various regimes in models with $U(1)\times U(1)$ symmetry  were considered as a function of magnetic-field penetration length and strength of mixed-gradient terms  in~\cite{babaev2002phase,babaev2004superconductor,Smorgrav2005_observation,Smiseth2005,kuklov2003counterflow,Kuklov2004,Kuklov2005,Kuklov2006,Dahl2008_preemptive,Herland2010,agterberg2008dislocations,berg2009charge,Grinenko2021_state,blomquist2021borromean}.
When such symmetry is present, the phase transition is driven by the proliferation of vortex loops in the two-component condensate (unless the system is strongly type-I).  These topological excitations can be denoted by a pair of two integers corresponding to the phase winding in each condensate:
 \begin{equation}
     (\Delta\phi_1=2 \pi n_1,\Delta\phi_2=2 \pi n_2 ) \equiv (n_1, n_2).
 \end{equation}

 Together with single-component vortices, which have phase windings in only one of the two condensates being $(1,0)$ or $(0,1)$, the system can also proliferate composite vortices   $(1,1)$ or   $(1, -1)$ leading to phases with composite order.

Various phases supported by the model can be seen more clearly by rewriting Eq.\eqref{continuum_London} in terms of charged and neutral modes~\cite{Babaev2002_hidden, Babaev2002_vortices, Herland2010,Smiseth2005,garaud2021skyrmions}: 

 \begin{equation}
\begin{split}
f= &\frac{1}{2}\Big\{ \left({\mathbf{\nabla}} \times \mathbf{A} \right)^2 +
\frac{\alpha - \nu^2}{\zeta} \left( {\mathbf{\nabla}} \phi_1 - {\mathbf{\nabla}} \phi_2\right)^2+ \\ &+ \frac{1}{\zeta} \left[  \left( 1 - \nu \right){\mathbf{\nabla}} \phi_1 +  \left( \alpha - \nu \right) {\mathbf{\nabla}} \phi_2  -\zeta e \mathbf{A} \right]^2 \Big\}+\\  &
+ \eta_2 \cos[2(\phi_1 -\phi_2)],
\label{continuum_London2}
\end{split}
\end{equation}

where:
\begin{equation}
    \zeta=  1 + \alpha - 2\nu.
    \label{m0}
\end{equation}
Consider first the case $\eta_2=0$. The most detailed Monte-Carlo calculations of the resulting phase diagram in three dimensions was presented in \cite{Herland2010}. As one can see from Eq.~\eqref{continuum_London2}, the gauge field is coupled with the phase-sum mode. As a consequence, for $e \neq 0$ the energetic cost per unit length of a $(1, 1)$ composite vortex is finite and can be made arbitrarily small by increasing the value of the electric charge. When $(1,1)$  vortices proliferate the system retains order in the phase difference.
That state is not a superconductor and has nontrivial magnetic properties,   described beyond the London limit by an effective model related to the Skyrme model \cite{garaud2021skyrmions}.
The energy cost of $(1, -1)$ in turn depends on the parameter $\nu$.
When $(1, -1)$ vortices proliferate, but $(1,0)$ or $(0,1)$ do not proliferate, the system retains order in the phase sum, representing a charge-$4e$ superconductor.

Let us now consider the case where $\eta_2 \neq 0$ and $\nu>0$.
The presence of an intercomponent biquadratic Josephson interaction changes the phase diagram of the model  by explicitly breaking the $U(1)\times U(1)$  symmetry down to a $U(1)\times Z_2$ symmetry, where the $U(1)$ symmetry is associated with the charged phase-sum mode. 

For $\eta_2 <0$ the phase difference $\phi_{1,2}=\phi_1 - \phi_2$ in the ground state can be either $0$ or $\pi$; while for  $\eta_2 >0$,  either $\pi/2$ or $-\pi/2$. In this work we will focus on the latter case, where the spontaneous breakdown of the $Z_2$ symmetry is associated with the spontaneous breaking of a time-reversal symmetry, since the complex conjugation of the order parameter leads to a different ground state.

The presence of a Josephson coupling also brings in a different type of topological excitations. These appear as domain walls separating in space two energetically equivalent states, namely a state where $\phi_{1,2}= \pi/2$ from another where  $\phi_{1,2}= -\pi/2$. So, just as the proliferation of composite-vortices of the kind $(1,1)$ will restore the $U(1)$ continuum symmetry, the proliferation of domain walls will restore the $Z_2$ time-reversal symmetry.
The phase diagram of the system will thus depend on the relative energetic cost of these two topological defects. 
In particular, a scenario where $T_c^{Z_2} > T_c$ can arise if the energetic cost of domain-walls nucleation is sufficiently high with respect that of 
$(1,1)$ vortices.
The parameters of the model, i.e. the gauge field coupling that parametrizes the magnetic field penetration length, the strength of the Josephson coupling, and the mixed-gradient coupling, each affect the relative cost of vortex and domain-wall excitations. However, the peculiarity of the model is such that simple energy arguments cannot be used to map out its phase diagram. Firstly, entropic factors are important and nontrivial. Secondly, domain walls are strongly and nontrivially interacting with vortices and have, under certain conditions, a tendency to form composite objects:  Skyrmions \cite{Garaud2011topological,Garaud2013chiral,garaud2014domain}. 
Hence we use Monte-Carlo approach to study the phase diagram of the system.

 \section{Details of the Monte Carlo simulations}

\subsection {Villain lattice model}

To perform Monte Carlo simulations, we need to provide a discrete lattice representation of the continuum model Eq.\eqref{continuum_London}. We consider a three-dimensional cubic lattice of size $L^3$ and lattice spacing $h=1$. The phases are defined on the lattice vertices $\phi_{{\bf{r}}, j}$ with $j=1,2$ labeling the two components. On the other hand, both the phase gradient, which is defined as the phase difference between two neighbouring sites $\partial_{\mu}\phi({\bf{r}})_j \to \Delta_{\mu} \phi_{{\bf{r}}, j}$, and the gauge field $A_{{\bf{r}}, \mu}$ are associated with the link connecting the vertex ${\bf r}$ with its neighbour ${\bf r} + \mu$, being $\mu = \hat{x}, \hat{y}$. The lattice curl of the gauge field, defined around a unitary plaquette, reads $(\nabla \times {\bf A (r)}) \to (\sum_{\kappa, \eta} \epsilon_{\mu \kappa \eta} \Delta_{\kappa} A_{{\bf{r}}, \eta})$, being $\epsilon_{\mu \kappa \eta}$ the Levi-Civita symbol.

As discussed in~\cite{Dahl2008_preemptive, Herland2010}, a good discretization scheme that allows an artifact-free representation of the dissipationless drag interaction is the multicomponent generalization of the Villain approximation~\cite{Villain1975}, which accommodates the compactness of the phase by rewriting:
\begin{equation*}
    e^{\beta \cos{ \left( \Delta_{\mu} \phi_i \right)} }  \to \sum_{n=-\infty}^{\infty} e^{-\frac{\beta}{2} (\phi_{i+\mu} - \phi_i- 2\pi n)^2},
\end{equation*}
where $\beta=1/T$ is the inverse temperature. The Villain Hamiltonian for the model~\eqref{continuum_London} reads:
\begin{equation}
\begin{split}
        H_v[\phi_1, \phi_2, {\bf{A}}; \beta] =  -\sum_{r, \mu} \beta^{-1} \ln \left\{ \sum_{n_{1,\mu} n_{2,\mu}= - \infty}^{\infty} e^{-{\beta} S} \right\} ,
        \label{H_Vill}
\end{split}
\end{equation}
where
\begin{equation}
\begin{split}
    S &= \frac{1}{2} [u_{{\bf{r}}, \mu, 1}^2 +\alpha u_{{\bf{r}}, \mu, 1}^2] - \nu ( u_{{\bf{r}}, \mu, 1} u_{{\bf{r}}, \mu, 2})+ \\&+ \frac{1}{2} (\sum_{\kappa, \eta} \epsilon_{\mu \kappa \eta} \Delta_{\kappa} A_{{\bf{r}}, \eta})^2 + \eta_2 \cos[2(\phi_{{\bf{r}}, 1} -\phi_{{\bf{r}}, 2} )], 
\end{split}
\end{equation}
and $u_{{\bf{r}}, \mu, j} = \Delta_{\mu} \phi_{{\bf{r}}, j} - e A_{{\bf{r}}, \mu} - 2\pi n_{{\bf{r}}, \mu, j} $. 

We have performed Monte Carlo (MC) simulations of the Villain Hamiltonian Eq.~\eqref{H_Vill}, locally updating the two phase fields $\phi_1, \phi_2 \in [0, 2\pi )$ as well as the gauge field ${\bf A}$ by means of the Metropolis-Hastings algorithm. A single MC step here consists of the Metropolis sweeps of the whole lattice fields. To speed-up the thermalization, we also implemented a parallel tempering algorithm, allowing swap of field configurations between neighbouring temperatures. Typically, we propose one set of swap after 32 MC steps.
For most of the numerical simulations, we performed a total of $2\times 10^5$ Monte Carlo steps, discarding the transient time occurring within the first $50000$ steps.
For the simulation performed in the limit of large Josephson coupling we implemented a cluster update to prevent the system from getting stuck in metastable states and we extended the total MC time up to $4\times 10^5$ steps discarding the first $150000$ steps. 
We have considered different values of the linear size $L$, as is needed to properly assess the critical points of the model.

\subsection{Locating the two phase transitions}

\subsubsection{The U(1) transition}

In the limiting case $e=0$, the $U(1)$ transition is associated with the onset of a superfluid phase, captured by the helicity modulus of the phase sum \cite{Dahl2008_preemptive, Herland2010}. 
The helicity modulus $\Upsilon$ measures the energetic cost associated with an infinitesimal twist of the order parameter phase across the system. 
In a multicomponent system, one can define several helicity moduli corresponding to different linear combinations of individual phase twist. In the two-component case, for each choice of the coefficients $\{ a_i \}$ in
\begin{equation}
\begin{pmatrix} \phi'_1(\mathbf{r}) \\  \phi'_2(\mathbf{r})  \end{pmatrix}
= \begin{pmatrix} \phi_1(\mathbf{r}) \\  \phi_2(\mathbf{r})  \end{pmatrix} 
+  \begin{pmatrix} a_1 \\  a_2  \end{pmatrix} \mathbf{\delta} \cdot \mathbf{r},
\end{equation}
one can define a corresponding helicity modulus
\begin{equation}
\begin{split}
\Upsilon_{\mu, \{a_i\}} &=\frac{1}{L^3} \frac{\partial^2 F(\{\phi'_i\}) }{\partial \delta_{\mu}^2}\Bigr|_{\delta_{\mu}=0}= \\&= \sum_{i} a_i^2 \Upsilon_{\mu, i} + 2\sum_{i<j} a_i a_j \Upsilon_{\mu, ij},
\end{split}
\label{Helicity_multicomponent1}
\end{equation}
where
\begin{equation}
\label{Helicity1}
\begin{split}
 \Upsilon_{\mu, i}= \frac{1}{L^3} \Big[ \Big\langle \frac{\partial^2 H}{ \partial \delta_{\mu,i}^2}
  \Big\rangle   -\beta \Big\langle \left(  \frac{\partial H}{ \partial \delta_{\mu,i}} - \langle \frac{\partial H}{ \partial \delta_{\mu,i}} \rangle \right)^2  \Big\rangle  \Big]_{\delta=0}; 
  \end{split}
 \end{equation}
 \begin{equation}
  \label{Helicity2}
 \begin{split}
 \Upsilon_{\mu, ij}=  \frac{1}{L^3} \Big[ &\Big\langle \frac{\partial^2 H}{ \partial \delta_{\mu,i} \partial \delta_{\mu,j} } \Big\rangle + \\&  - \beta \Big\langle \left(  \frac{\partial H}{ \partial \delta_{\mu,i}} - \langle \frac{\partial H}{ \partial \delta_{\mu,j}} \rangle \right)^2  \Big\rangle \Big]_{\delta=0};
 \end{split}
\end{equation}
and $\langle \dots \rangle$ stays for the thermal average over the MC steps.
Since the $U(1)$ superfluid phase transition is associated with the phase-sum mode, the relevant observable is the phase-sum helicity modulus $\Upsilon_{+}$ defined by the choice $a_1 = a_2 = 1$.
\begin{figure*}[t!]
    \centering
    \includegraphics[width=\linewidth]{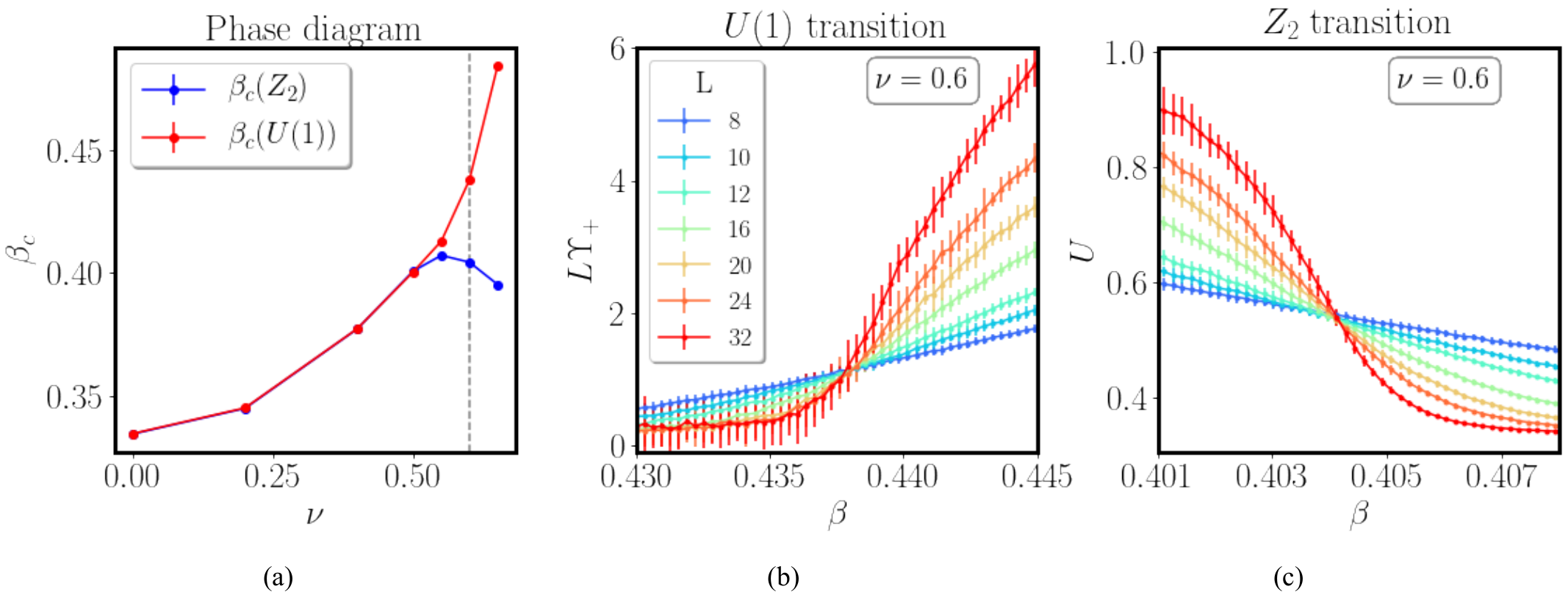}
    \caption{(a) Phase diagram of the model \eqref{continuum_London}  as a function of interaction coupling $\nu$ for the extreme type-II limit ($e=0$) with Josephson coupling fixed to $\eta_2=0.1$. The $U(1)$ inverse critical temperature clearly splits apart from the $Z_2$ critical temperature for $\nu > 0.5$, where the BTRS quartic metal phase arises.   
    The gray dashed line indicates the value $\nu=0.6$.
    (b) Helicity-modulus sum $\Upsilon_+$ for $\nu=0.6$ rescaled by the linear size of the system $L$ as function of the inverse temperature $\beta$. Different values of $L$ are shown.  (c)  Binder cumulant $U$ for $\nu=0.6$ as function of the inverse temperature $\beta$, for different values of the linear size $L$. }
    \label{fig1}
\end{figure*}
We computed $\Upsilon_{+}$ for different values of the linear sizes $L$ and determined the critical temperature $T_c$ by means of the finite-size crossings, extrapolated to the thermodynamic limit, of the quantity $L \Upsilon_{+}$.

In the case where $e \neq 0$, we locates the $U(1)$ superconducting transition by computing the dual stiffness \cite{Motrunich2008, Herland2013, Carlstrom2015, Grinenko2021_state}, that accounts for the onset of the SC state by measuring the Meissner effect:

\begin{equation}
  \rho^\mu(\mathbf{q}) =
    \left\langle \frac{\left| \sum_{\mathbf{r},\nu,\lambda}
      \epsilon_{\mu\nu\lambda} \left[\Delta_\nu A_\lambda(\mathbf{r})\right]
      \mathrm{e}^{\mathrm{i}\mathbf{q} \cdot \mathbf{r}} \right|^2}
      {(2\pi)^2 L^3} \right\rangle.
\end{equation}

We compute the dual stiffness in the $z$ direction at the smallest relevant wave vector in the $x$ direction, i.e. $\mathbf{q}_\mathrm{min}^x = (2\pi/L, 0, 0)$. In what follows, we denote $\rho^z(\mathbf{q}_\mathrm{min}^x)$ simply as $\rho$.  This observable accounts for the long-range fluctuations of the magnetic field, suppressed in the superconducting phase and finite in the normal phase. Thus, contrarily to the superfluid stiffness, it is expected to be zero in the superconducting phase and non-zero in the normal one. 
Finally, we use finite-size crossings of $L\rho$, extrapolated to the thermodynamic limit, in order to locate superconducting transitions.

\subsubsection{The $Z_2$ phase transition}

We define a $Z_2$ Ising order parameter $m$ associated with the two possible degeneracy of the ground state. In particular, we set $m$ to be equal to $+1 $ or $-1$ according with the sign of the phase difference $\phi_{1,2}$.

In order to locate the $Z_2$ critical temperature, we compute the Binder cumulant $U$~\cite{Binder1981_critical, Binder1981_finitesize} for the order parameter $m$:

\begin{equation}
  U = \frac{\langle m^4 \rangle}{3\langle m^2 \rangle^2},
  \label{binder_cumulant}
\end{equation}
which is expected to be a universal quantity at the critical point.

The critical temperature $T_c^{Z_2}$, associated with the spontaneous breaking of the time-reversal symmetry, is thus determined by means of finite-size crossings of the Binder cumulant $U$ extrapolated to the thermodynamic limit.

The error bars of all the observables are estimated via a bootstrap resampling method. In the figures shown, when not visible, the estimated error bars are smaller than the symbol sizes.

\section{Results}
Here we present the Monte Carlo numerical results obtained for different regimes of the two-component London model \eqref{continuum_London}.

\subsection{Extreme type-II regime $e=0$: the equal densities case $\alpha=1$.}

Let us start discussing the limiting case of infinite penetration length $\lambda \to \infty$, i.e. $e=0$. In the isotropic case where $\alpha=1$, the free energy \eqref{continuum_London} reads:

\begin{equation}
\begin{split}
    f= & \frac{1+\nu}{4}\left( {\mathbf{\nabla}} \phi_1 -{\mathbf{\nabla}} \phi_2 \right)^2 + \frac{1-\nu}{4}\left( {\mathbf{\nabla}} \phi_1 +{\mathbf{\nabla}} \phi_2 \right)^2+\\& +  \eta_2 \cos[2(\phi_1  -\phi_2  )],
    \end{split}
    \label{e0_iso_continuum_London_chargeneutral}
\end{equation}
where we fixed $\eta_2=0.1$. 

The increase of the drag coupling $\nu$ reduces the energetic cost of nucleating $(1,1)$ composite vortices, whose proliferation restores the superconducting $U(1)$ symmetry, with respect to that domain walls, associated with the  the $Z_2$ time-reversal symmetry.

The two inverse critical temperatures, respectively $\beta_c(U(1))$ and $\beta_c(Z_2)$ are reported in Fig.\ref{fig1}(a) as function of the coupling $\nu$. For low values of $\nu$, the simulations indicate a single phase transition dividing a low-temperature superconducting phase, where both the $U(1)$ and the $Z_2$ time-reversal symmetry are spontaneously broken, from a high-temperature phase where the system is in its normal metal phase.
On the other hand, for sufficiently large $\nu$, the two phase transitions split apart with $\beta_c(U(1)) > \beta_c(Z_2)$. In this regime, the system shows a fermionic-quandrupling phase arising between the low-temperature superconducting phase and the high-temperature normal phase. Such a state is disordered in the phase-sum mode, i.e. it is not superconducting, but still ordered in the phase-difference mode due to the spontaneous breaking of the $Z_2$ time-reversal symmetry. 
The order parameter characterizing this phase is thus the two-component phase difference, fourth order in terms of fermionic fields being:  $\phi_{1,2}= \arccos{\frac{1}{2}[\psi_1\psi_2^* + \psi_1^*\psi_2]}$.
The region in the phase diagram Fig.\ref{fig1}(a) where this BTRS quartic metal phase arises becomes wider for higher values of $\nu$.

\subsection{Extreme type-II regime $e=0$: the effects of component disparity $\alpha <1$}

Next, we consider the case where the two condensates have different densities $\rho_1 \neq \rho_2$, i.e. $\alpha \neq 1$. We choose two different values of the dissipationless drag coupling constant so to consider both the two regimes:  $\nu=0.4$  the one where $\beta_c(U(1)) = \beta_c(Z_2)$ and $\nu=0.6$ where $\beta_c(U(1)) > \beta_c(Z_2)$. 

In the present case the free energy reads:

\begin{equation}
\begin{split}
    f= & \frac{\alpha - \nu^2}{\zeta}\left( {\mathbf{\nabla}} \phi_1 -{\mathbf{\nabla}} \phi_2 \right)^2 +\\ & + \frac{1}{\zeta}\left[ (1-\nu) {\mathbf{\nabla}} \phi_1 +(\alpha-\nu){\mathbf{\nabla}} \phi_2 \right]^2+\\& +  \eta_2 \cos[2(\phi_1  -\phi_2  )],
    \end{split}
    \label{e0_aniso_continuum_London_chargeneutral}
\end{equation}
with $\zeta=1+\alpha -2 \nu$.
Since here we are studying the model for fixed values of $\nu$ as function of $\alpha$ it is convenient to rewrite the stability condition of Eq.\eqref{e0_aniso_continuum_London_chargeneutral} as: $\alpha > \nu$, which automatically fulfill $\alpha > \nu^2$, being $\nu<1$.
\begin{figure}[h!]
    \centering
    \includegraphics[width=0.88\linewidth]{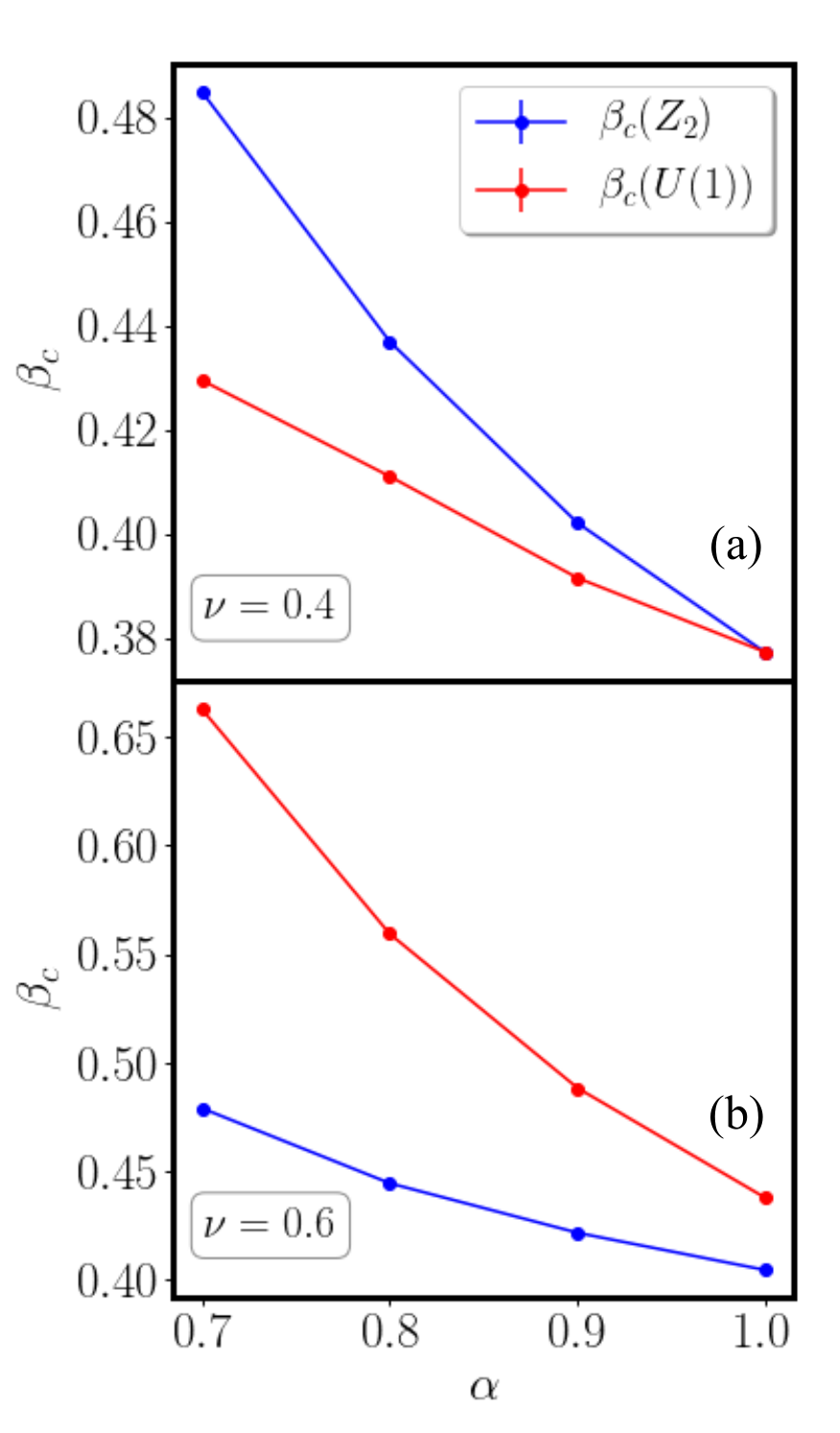}
    \caption{$U(1)$ and $Z_2$ inverse critical temperatures as a function of the  component disparity parameter $\alpha$. The results shown have been obtained for $e=0$, $\eta_2=0.1$ and respectively: (a) $\nu=0.4$ and (b) $\nu=0.6$. }
    \label{fig2}
\end{figure}

The Monte Carlo numerical results for the model \eqref{e0_aniso_continuum_London_chargeneutral} are shown in Fig.\ref{fig2}(a)-(b) for $\nu=0.4$ and  for $\nu=0.6$ respectively.
In both cases, decreasing $\alpha$, the two transitions move to lower critical temperatures. However, for $\nu=0.4$ (Fig.\ref{fig2}(a)),  we find that the inverse critical temperature associated with the time-reversal symmetry breaking $\beta_c(Z_2)$ grows faster than the one associated to the $U(1)$ superconducting symmetry $\beta_c(U(1))$, with a resulting split of the two transitions for $\alpha <1$. In this case, the split occurs with $\beta_c(Z_2)> \beta_c(U(1)) $, implying a relative increase of the time-reversal invariant superconducting state.
On the other hand, for the case $\nu=0.6$ (Fig.\ref{fig2}(b)), we find the opposite scenario: by decreasing $\alpha$ the inverse critical temperature $\beta_c(U(1))$ increases faster than $\beta_c(Z_2)$.

\begin{figure*}[t!]
    \centering
    \includegraphics[width=\linewidth]{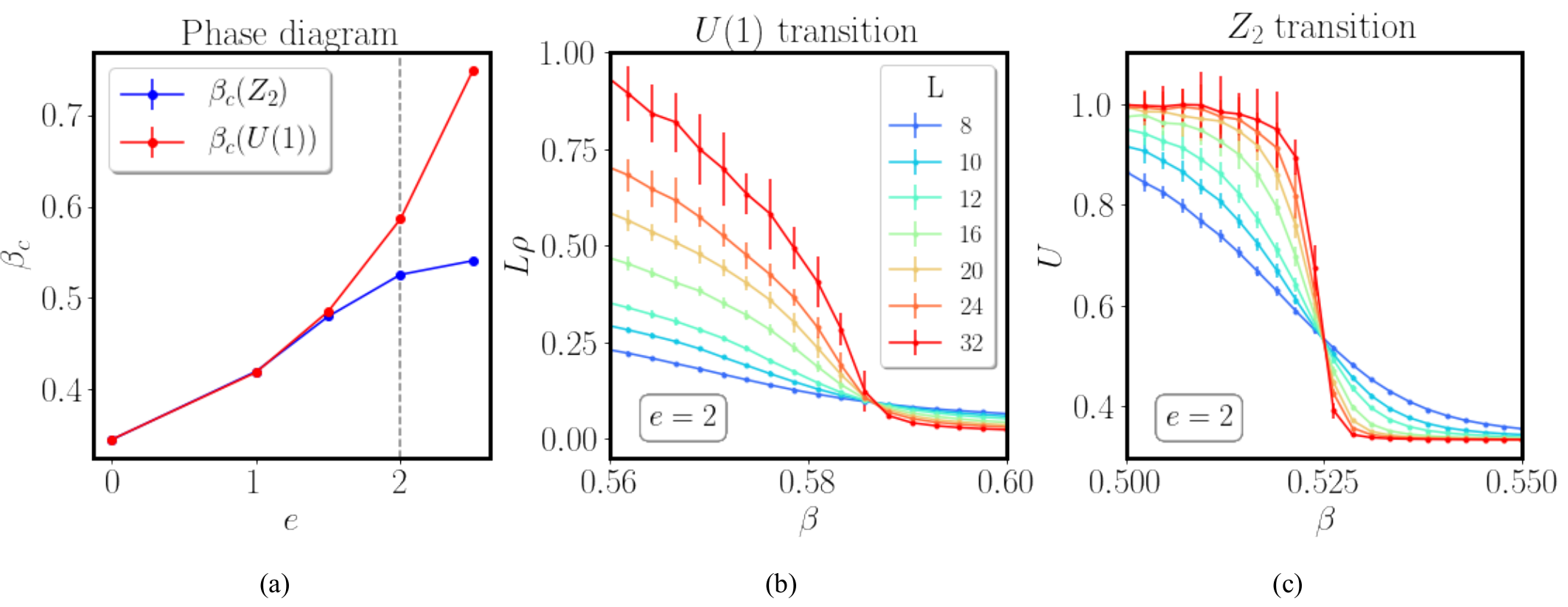}
    \caption{(a) Phase diagram as function of the electric charge $e$ (which parametrizes the magnetic field penetration length) for the mixed-gradient coupling $\nu=0.2$. The $U(1)$ inverse critical temperature split apart from the $Z_2$ for $e \geq 1.5$, where the BTRS quartic metal phase arises.   
    The gray dashed line indicates the value $e=2$.
    (b) Dual stiffness $\rho$ for $e=2$ rescaled by the linear size of the system $L$ as function of the inverse temperature $\beta$. Different values of $L$ are shown.  (c)  Binder cumulant $U$ for $e=2$ as function of the inverse temperature $\beta$, for different values of the linear size $L$. }
    \label{fig3}
\end{figure*}
\subsection{Charged case $e\neq0$}

After presenting the numerical results obtained for the extreme type-II limit, let us now consider the phase diagram as a function of magnetic field penetration length relative to other length scales.  We consider two equivalent superconducting components $\rho_1=\rho_2$, i.e. $\alpha=1$. 

The free energy of the system reads: 

 \begin{equation}
\begin{split}
f= &\frac{1}{2}\Big\{ \left({\mathbf{\nabla}} \times \mathbf{A} \right)^2 +
\frac{1 + \nu}{2} \left( {\mathbf{\nabla}} \phi_1 - {\mathbf{\nabla}} \phi_2\right)^2+ \\ &+ \frac{1- \nu}{2} \left( {\mathbf{\nabla}} \phi_1 + {\mathbf{\nabla}} \phi_2  - 2 e \mathbf{A} \right)^2 \Big\}+\\  &
+ \eta_2 \cos[2(\phi_1 -\phi_2)].
\label{iso_charged_continuum_London2}
\end{split}
\end{equation}
We fixed the value of the Josephson coupling  $\eta_2=0.1$ and the dissipationless drag interaction to $\nu=0.2$, studying Eq.\eqref{iso_charged_continuum_London2} for different values of the electric charge $e$. The resulting phase diagram is shown in Fig.\ref{fig3} (a). 
For small values of $e$, the phase diagram for $\nu=0.2$ does not qualitatively change with respect to the neutral case: we can resolve only one phase transition associated with the spontaneous breaking of the total $U(1) \times Z_2 $ symmetry.  
\begin{figure}[b!]
    \centering
    \includegraphics[width=0.88\linewidth]{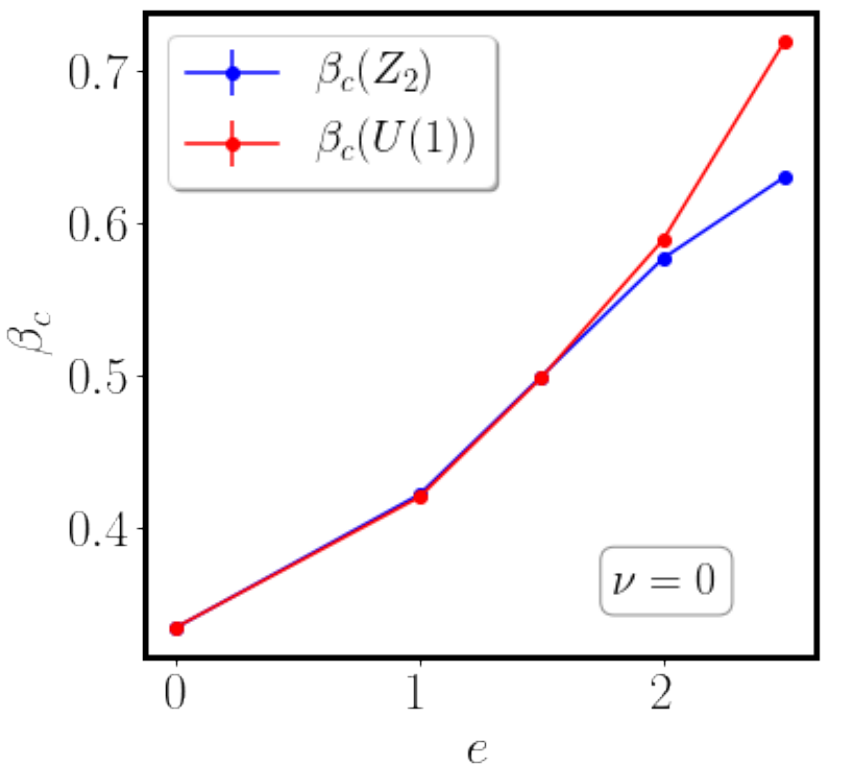}
    \caption{Phase diagram of the model \eqref{iso_charged_continuum_London2} as function of the electric charge $e$ for fixed Josephson coupling $\eta_2=0.1$ and in the limiting case where $\nu=0$. Without dissipationless drag interaction, the BTRS quartic metal phase is observed for $e > 1.5$.  }
    \label{fig4}
\end{figure}

However, by further increasing $e$, the energy of $(1,1)$ vortex-loops decreases. This reduces the critical temperature associated with the $U(1)$ symmetry breaking, without equally affecting the $Z_2$ order. As a result, the two phase transitions separate for large enough $e$, with the appearance of the BTRS quartic metallic phase.

Consistently with  the results obtained for $\nu=0$ in a different discretization scheme of the London model  ~\cite{Bojesen2014_phase},  we obtain that for strong enough coupling with the gauge field, the splitting of the two phase transitions  appears in this model even for $\nu=0$ 
Fig.\ref{fig4}.

\subsection{Large Josephson coupling $\eta_2$ limit }
 
 We finally consider the limiting case of very large Josephson coupling $\eta_2$ at fixed values of the electric charge $e$ and the dissipationless drag interaction $\nu$. 
 As $\eta_2$ increases, the energy cost of nucleating a domain wall increases, with a resulting increase of the critical temperature associated with the $Z_2$ time-reversal symmetry breaking. 
 However, the Josephson coupling  is a local coupling, hence for very large values of $\eta_2$ the energy cost of a domain wall will saturate since the domain-wall width cannot be larger than the lattice spacing. Thus one would expect the saturation of   $T_c(Z_2)$ in the lattice London model.
 
  \begin{figure}[h!]
    \centering
    \includegraphics[width=0.88\linewidth]{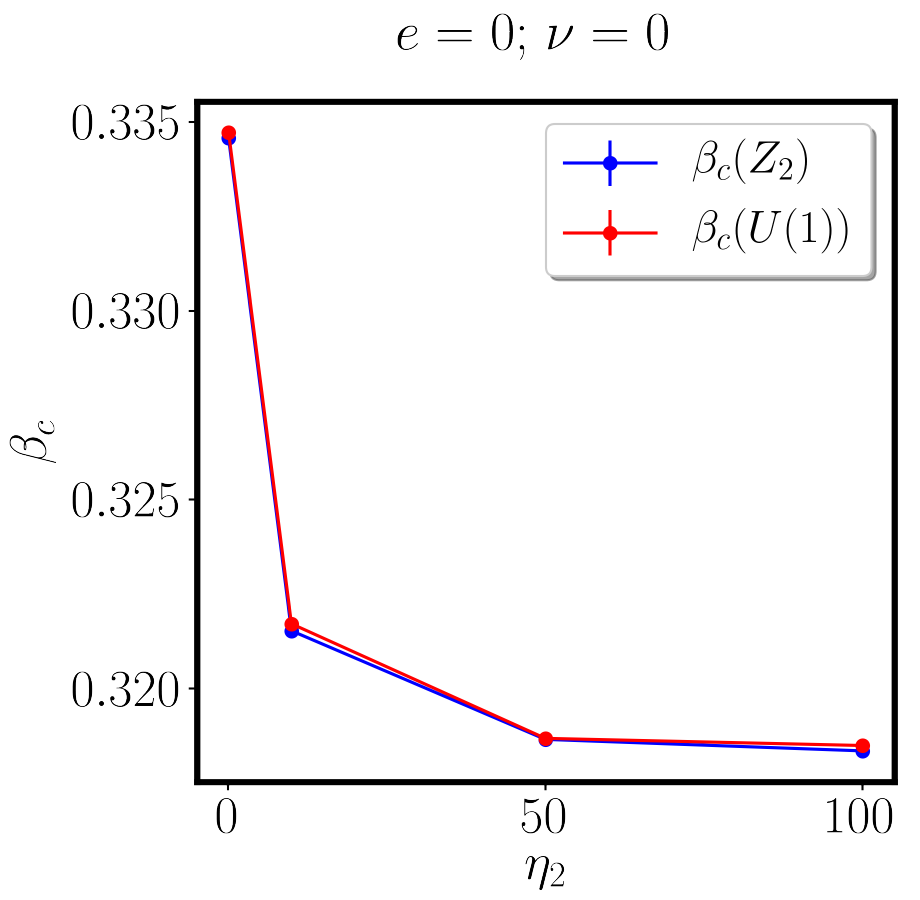}
    \caption{Phase diagram of the model \eqref{e0_iso_continuum_London_chargeneutral} for $e=0$ and $\nu=0$ as function of the Josephson coupling $\eta_2$.}
    \label{fig6}
\end{figure}
 
 As reported in Fig.\ref{fig6}, for the model \eqref{continuum_London} with $\nu=0$ and $e=0$ in our numerical results we cannot resolve the splitting for the $Z_2$ and $U(1)$ phase transitions, at least for the system sizes that we simulated.
 
\begin{figure}[h!]
    \centering
    \includegraphics[width=0.88\linewidth]{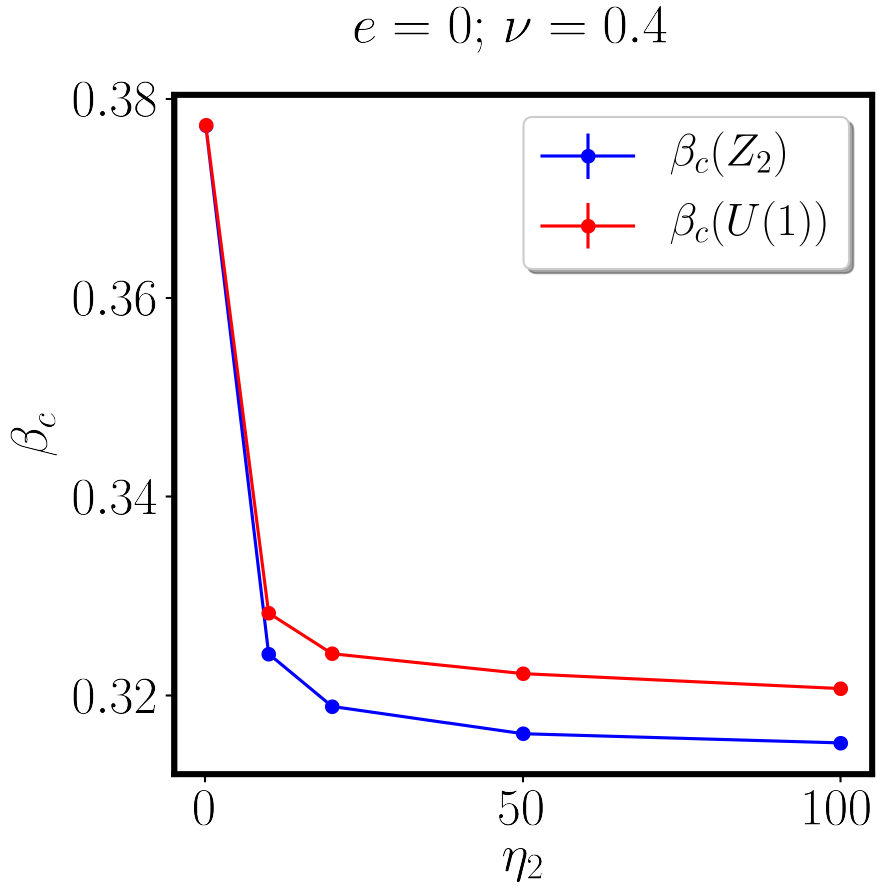}
    \caption{Phase diagram of the model \eqref{e0_iso_continuum_London_chargeneutral} for $e=0$ and $\nu=0.4$ as function of the Josephson coupling $\eta_2$. For $\eta_2 \geq 10$ the two phase transitions split apart with $\beta_c(Z_2)> \beta_c(U(1))$. In such temperature range between  $\beta_c(U(1))$ and $\beta_c(Z_2)$ there is a BTRS quartic metallic phase.  }
    \label{fig7}
\end{figure}

 Nevertheless, large values of the Josephson coupling $\eta_2$ widen the region of the phase diagram Fig.\ref{fig1}(a) where the BTRS quartic metal phase appears. For $\nu=0.4$, as reported in Fig.\ref{fig7}, the two phase transitions separate significantly for $\eta_2=1$.

\section{Conclusions}
In conclusion, the pairwise interaction of electrons usually leads to the formation of pair condensates. However, in systems that break multiple symmetries, there could be fluctuations-induced  states corresponding to fermionic-quadrupling condensates.
These states appear from the proliferation of bound states of topological defects that partially restore symmetry and lead to an effective fermionic-quadrupling interaction.\\

We studied the appearance of a fermionic-quadrupling condensate that breaks time-reversal symmetry, as reported in a recent experimental work~\cite{Grinenko2021_state} in Ba$_{1-x}$K$_x$Fe$_2$As$_2$ with doping level $x\approx 0.8$. The phase shows order only in the phase differences between the components of complex fields.  While the precise microscopic model for this compound is not established yet, we
  studied the appearance of such a state in a three-dimensional two-component London model.
We  presented the phase diagram of the model and appearance of the quartic state as a function of the magnetic field penetration length (parameterized by the gauge field coupling constant), the strength of biquadratic Josephson, and mixed-gradient couplings.

\section{Acknowledgements}
We thank Daniel Weston, Vadim Grinenko, Egil Herland, and David Aceituno for many discussions.
The simulations were performed on resources provided by the Swedish
National Infrastructure for Computing (SNIC) at the National Supercomputer Center
at Link\"oping, Sweden. I.M. acknowledges the Carl Trygger foundation
through grant number CTS 20:75. E.B. is supported by the Swedish Research Council Grants  2016-06122, 2018-03659.

\bibliography{weston}

\begin{thebibliography}{58}%
\makeatletter
\providecommand \@ifxundefined [1]{%
 \@ifx{#1\undefined}
}%
\providecommand \@ifnum [1]{%
 \ifnum #1\expandafter \@firstoftwo
 \else \expandafter \@secondoftwo
 \fi
}%
\providecommand \@ifx [1]{%
 \ifx #1\expandafter \@firstoftwo
 \else \expandafter \@secondoftwo
 \fi
}%
\providecommand \natexlab [1]{#1}%
\providecommand \enquote  [1]{``#1''}%
\providecommand \bibnamefont  [1]{#1}%
\providecommand \bibfnamefont [1]{#1}%
\providecommand \citenamefont [1]{#1}%
\providecommand \href@noop [0]{\@secondoftwo}%
\providecommand \href [0]{\begingroup \@sanitize@url \@href}%
\providecommand \@href[1]{\@@startlink{#1}\@@href}%
\providecommand \@@href[1]{\endgroup#1\@@endlink}%
\providecommand \@sanitize@url [0]{\catcode `\\12\catcode `\$12\catcode
  `\&12\catcode `\#12\catcode `\^12\catcode `\_12\catcode `\%12\relax}%
\providecommand \@@startlink[1]{}%
\providecommand \@@endlink[0]{}%
\providecommand \url  [0]{\begingroup\@sanitize@url \@url }%
\providecommand \@url [1]{\endgroup\@href {#1}{\urlprefix }}%
\providecommand \urlprefix  [0]{URL }%
\providecommand \Eprint [0]{\href }%
\providecommand \doibase [0]{http://dx.doi.org/}%
\providecommand \selectlanguage [0]{\@gobble}%
\providecommand \bibinfo  [0]{\@secondoftwo}%
\providecommand \bibfield  [0]{\@secondoftwo}%
\providecommand \translation [1]{[#1]}%
\providecommand \BibitemOpen [0]{}%
\providecommand \bibitemStop [0]{}%
\providecommand \bibitemNoStop [0]{.\EOS\space}%
\providecommand \EOS [0]{\spacefactor3000\relax}%
\providecommand \BibitemShut  [1]{\csname bibitem#1\endcsname}%
\let\auto@bib@innerbib\@empty
\bibitem [{\citenamefont {Bardeen}\ \emph
  {et~al.}(1957{\natexlab{a}})\citenamefont {Bardeen}, \citenamefont {Cooper},\
  and\ \citenamefont {Schrieffer}}]{Bardeen1957_microscopic}%
  \BibitemOpen
  \bibfield  {author} {\bibinfo {author} {\bibfnamefont {J.}~\bibnamefont
  {Bardeen}}, \bibinfo {author} {\bibfnamefont {L.~N.}\ \bibnamefont {Cooper}},
  \ and\ \bibinfo {author} {\bibfnamefont {J.~R.}\ \bibnamefont {Schrieffer}},\
  }\href {\doibase 10.1103/PhysRev.106.162} {\bibfield  {journal} {\bibinfo
  {journal} {Phys. Rev.}\ }\textbf {\bibinfo {volume} {106}},\ \bibinfo {pages}
  {162} (\bibinfo {year} {1957}{\natexlab{a}})}\BibitemShut {NoStop}%
\bibitem [{\citenamefont {Bardeen}\ \emph
  {et~al.}(1957{\natexlab{b}})\citenamefont {Bardeen}, \citenamefont {Cooper},\
  and\ \citenamefont {Schrieffer}}]{Bardeen1957_theory}%
  \BibitemOpen
  \bibfield  {author} {\bibinfo {author} {\bibfnamefont {J.}~\bibnamefont
  {Bardeen}}, \bibinfo {author} {\bibfnamefont {L.~N.}\ \bibnamefont {Cooper}},
  \ and\ \bibinfo {author} {\bibfnamefont {J.~R.}\ \bibnamefont {Schrieffer}},\
  }\href {\doibase 10.1103/PhysRev.108.1175} {\bibfield  {journal} {\bibinfo
  {journal} {Phys. Rev.}\ }\textbf {\bibinfo {volume} {108}},\ \bibinfo {pages}
  {1175} (\bibinfo {year} {1957}{\natexlab{b}})}\BibitemShut {NoStop}%
\bibitem [{\citenamefont {Babaev}(2002{\natexlab{a}})}]{babaev2002phase}%
  \BibitemOpen
  \bibfield  {author} {\bibinfo {author} {\bibfnamefont {E.}~\bibnamefont
  {Babaev}},\ }\href@noop {} {\bibfield  {journal} {\bibinfo  {journal} {arXiv
  preprint cond-mat/0201547}\ } (\bibinfo {year}
  {2002}{\natexlab{a}})}\BibitemShut {NoStop}%
\bibitem [{\citenamefont {Babaev}\ \emph {et~al.}(2004)\citenamefont {Babaev},
  \citenamefont {Sudb\"o},\ and\ \citenamefont
  {Ashcroft}}]{babaev2004superconductor}%
  \BibitemOpen
  \bibfield  {author} {\bibinfo {author} {\bibfnamefont {E.}~\bibnamefont
  {Babaev}}, \bibinfo {author} {\bibfnamefont {A.}~\bibnamefont {Sudb\"o}}, \
  and\ \bibinfo {author} {\bibfnamefont {N.}~\bibnamefont {Ashcroft}},\
  }\href@noop {} {\bibfield  {journal} {\bibinfo  {journal} {Nature}\ }\textbf
  {\bibinfo {volume} {431}},\ \bibinfo {pages} {666} (\bibinfo {year}
  {2004})}\BibitemShut {NoStop}%
\bibitem [{\citenamefont {Sm\"orgrav}\ \emph {et~al.}(2005)\citenamefont
  {Sm\"orgrav}, \citenamefont {Babaev}, \citenamefont {Smiseth},\ and\
  \citenamefont {Sudb\"o}}]{Smorgrav2005_observation}%
  \BibitemOpen
  \bibfield  {author} {\bibinfo {author} {\bibfnamefont {E.}~\bibnamefont
  {Sm\"orgrav}}, \bibinfo {author} {\bibfnamefont {E.}~\bibnamefont {Babaev}},
  \bibinfo {author} {\bibfnamefont {J.}~\bibnamefont {Smiseth}}, \ and\
  \bibinfo {author} {\bibfnamefont {A.}~\bibnamefont {Sudb\"o}},\ }\href
  {\doibase 10.1103/PhysRevLett.95.135301} {\bibfield  {journal} {\bibinfo
  {journal} {Phys. Rev. Lett.}\ }\textbf {\bibinfo {volume} {95}},\ \bibinfo
  {pages} {135301} (\bibinfo {year} {2005})}\BibitemShut {NoStop}%
\bibitem [{\citenamefont {Bojesen}\ \emph {et~al.}(2013)\citenamefont
  {Bojesen}, \citenamefont {Babaev},\ and\ \citenamefont
  {Sudb\"o}}]{Bojesen2013time}%
  \BibitemOpen
  \bibfield  {author} {\bibinfo {author} {\bibfnamefont {T.~A.}\ \bibnamefont
  {Bojesen}}, \bibinfo {author} {\bibfnamefont {E.}~\bibnamefont {Babaev}}, \
  and\ \bibinfo {author} {\bibfnamefont {A.}~\bibnamefont {Sudb\"o}},\ }\href
  {\doibase 10.1103/PhysRevB.88.220511} {\bibfield  {journal} {\bibinfo
  {journal} {Phys. Rev. B}\ }\textbf {\bibinfo {volume} {88}},\ \bibinfo
  {pages} {220511 (R)} (\bibinfo {year} {2013})}\BibitemShut {NoStop}%
\bibitem [{\citenamefont {Bojesen}\ \emph {et~al.}(2014)\citenamefont
  {Bojesen}, \citenamefont {Babaev},\ and\ \citenamefont
  {Sudb\"o}}]{Bojesen2014_phase}%
  \BibitemOpen
  \bibfield  {author} {\bibinfo {author} {\bibfnamefont {T.~A.}\ \bibnamefont
  {Bojesen}}, \bibinfo {author} {\bibfnamefont {E.}~\bibnamefont {Babaev}}, \
  and\ \bibinfo {author} {\bibfnamefont {A.}~\bibnamefont {Sudb\"o}},\ }\href
  {\doibase 10.1103/PhysRevB.89.104509} {\bibfield  {journal} {\bibinfo
  {journal} {Phys. Rev. B}\ }\textbf {\bibinfo {volume} {89}},\ \bibinfo
  {pages} {104509} (\bibinfo {year} {2014})}\BibitemShut {NoStop}%
\bibitem [{\citenamefont {Agterberg}\ and\ \citenamefont
  {Tsunetsugu}(2008)}]{agterberg2008dislocations}%
  \BibitemOpen
  \bibfield  {author} {\bibinfo {author} {\bibfnamefont {D.}~\bibnamefont
  {Agterberg}}\ and\ \bibinfo {author} {\bibfnamefont {H.}~\bibnamefont
  {Tsunetsugu}},\ }\href@noop {} {\bibfield  {journal} {\bibinfo  {journal}
  {Nature Physics}\ }\textbf {\bibinfo {volume} {4}},\ \bibinfo {pages} {639}
  (\bibinfo {year} {2008})}\BibitemShut {NoStop}%
\bibitem [{\citenamefont {Berg}\ \emph {et~al.}(2009)\citenamefont {Berg},
  \citenamefont {Fradkin},\ and\ \citenamefont {Kivelson}}]{berg2009charge}%
  \BibitemOpen
  \bibfield  {author} {\bibinfo {author} {\bibfnamefont {E.}~\bibnamefont
  {Berg}}, \bibinfo {author} {\bibfnamefont {E.}~\bibnamefont {Fradkin}}, \
  and\ \bibinfo {author} {\bibfnamefont {S.~A.}\ \bibnamefont {Kivelson}},\
  }\href@noop {} {\bibfield  {journal} {\bibinfo  {journal} {Nature Physics}\
  }\textbf {\bibinfo {volume} {5}},\ \bibinfo {pages} {830} (\bibinfo {year}
  {2009})}\BibitemShut {NoStop}%
\bibitem [{\citenamefont {Herland}\ \emph {et~al.}(2010)\citenamefont
  {Herland}, \citenamefont {Babaev},\ and\ \citenamefont
  {Sudb\"o}}]{Herland2010}%
  \BibitemOpen
  \bibfield  {author} {\bibinfo {author} {\bibfnamefont {E.~V.}\ \bibnamefont
  {Herland}}, \bibinfo {author} {\bibfnamefont {E.}~\bibnamefont {Babaev}}, \
  and\ \bibinfo {author} {\bibfnamefont {A.}~\bibnamefont {Sudb\"o}},\ }\href
  {\doibase 10.1103/PhysRevB.82.134511} {\bibfield  {journal} {\bibinfo
  {journal} {Phys. Rev. B}\ }\textbf {\bibinfo {volume} {82}},\ \bibinfo
  {pages} {134511} (\bibinfo {year} {2010})}\BibitemShut {NoStop}%
\bibitem [{\citenamefont {Cho}\ \emph {et~al.}(2020)\citenamefont {Cho},
  \citenamefont {Shen}, \citenamefont {Lyu}, \citenamefont {Atanov},
  \citenamefont {Chen}, \citenamefont {Lee}, \citenamefont {San~Hor},
  \citenamefont {Gawryluk}, \citenamefont {Pomjakushina}, \citenamefont
  {Bartkowiak}, \citenamefont {Hecker}, \citenamefont {Schmalian},\ and\
  \citenamefont {Lortz}}]{cho2020z}%
  \BibitemOpen
  \bibfield  {author} {\bibinfo {author} {\bibfnamefont {C.-w.}\ \bibnamefont
  {Cho}}, \bibinfo {author} {\bibfnamefont {J.}~\bibnamefont {Shen}}, \bibinfo
  {author} {\bibfnamefont {J.}~\bibnamefont {Lyu}}, \bibinfo {author}
  {\bibfnamefont {O.}~\bibnamefont {Atanov}}, \bibinfo {author} {\bibfnamefont
  {Q.}~\bibnamefont {Chen}}, \bibinfo {author} {\bibfnamefont {S.~H.}\
  \bibnamefont {Lee}}, \bibinfo {author} {\bibfnamefont {Y.}~\bibnamefont
  {San~Hor}}, \bibinfo {author} {\bibfnamefont {D.~J.}\ \bibnamefont
  {Gawryluk}}, \bibinfo {author} {\bibfnamefont {E.}~\bibnamefont
  {Pomjakushina}}, \bibinfo {author} {\bibfnamefont {M.}~\bibnamefont
  {Bartkowiak}}, \bibinfo {author} {\bibfnamefont {M.}~\bibnamefont {Hecker}},
  \bibinfo {author} {\bibfnamefont {J.}~\bibnamefont {Schmalian}}, \ and\
  \bibinfo {author} {\bibfnamefont {R.}~\bibnamefont {Lortz}},\ }\href@noop {}
  {\bibfield  {journal} {\bibinfo  {journal} {Nature communications}\ }\textbf
  {\bibinfo {volume} {11}},\ \bibinfo {pages} {1} (\bibinfo {year}
  {2020})}\BibitemShut {NoStop}%
\bibitem [{\citenamefont {Fischer}\ and\ \citenamefont
  {Berg}(2016)}]{fischer2016fluctuation}%
  \BibitemOpen
  \bibfield  {author} {\bibinfo {author} {\bibfnamefont {M.~H.}\ \bibnamefont
  {Fischer}}\ and\ \bibinfo {author} {\bibfnamefont {E.}~\bibnamefont {Berg}},\
  }\href@noop {} {\bibfield  {journal} {\bibinfo  {journal} {Physical Review
  B}\ }\textbf {\bibinfo {volume} {93}},\ \bibinfo {pages} {054501} (\bibinfo
  {year} {2016})}\BibitemShut {NoStop}%
\bibitem [{\citenamefont {Shaffer}\ \emph {et~al.}(2021)\citenamefont
  {Shaffer}, \citenamefont {Wang},\ and\ \citenamefont
  {Santos}}]{shaffer2021theory}%
  \BibitemOpen
  \bibfield  {author} {\bibinfo {author} {\bibfnamefont {D.}~\bibnamefont
  {Shaffer}}, \bibinfo {author} {\bibfnamefont {J.}~\bibnamefont {Wang}}, \
  and\ \bibinfo {author} {\bibfnamefont {L.~H.}\ \bibnamefont {Santos}},\
  }\href@noop {} {\bibfield  {journal} {\bibinfo  {journal} {Physical Review
  B}\ }\textbf {\bibinfo {volume} {104}},\ \bibinfo {pages} {184501} (\bibinfo
  {year} {2021})}\BibitemShut {NoStop}%
\bibitem [{\citenamefont {Chung}\ and\ \citenamefont
  {Kim}(2022)}]{Chung2022berezinskii}%
  \BibitemOpen
  \bibfield  {author} {\bibinfo {author} {\bibfnamefont {S.~B.}\ \bibnamefont
  {Chung}}\ and\ \bibinfo {author} {\bibfnamefont {S.~K.}\ \bibnamefont
  {Kim}},\ }\href {\doibase 10.21468/SciPostPhysCore.5.1.003} {\bibfield
  {journal} {\bibinfo  {journal} {SciPost Phys. Core}\ }\textbf {\bibinfo
  {volume} {5}},\ \bibinfo {pages} {3} (\bibinfo {year} {2022})}\BibitemShut
  {NoStop}%
\bibitem [{\citenamefont {Kuklov}\ and\ \citenamefont
  {Svistunov}(2003)}]{kuklov2003counterflow}%
  \BibitemOpen
  \bibfield  {author} {\bibinfo {author} {\bibfnamefont {A.}~\bibnamefont
  {Kuklov}}\ and\ \bibinfo {author} {\bibfnamefont {B.~V.}\ \bibnamefont
  {Svistunov}},\ }\href@noop {} {\bibfield  {journal} {\bibinfo  {journal}
  {Phys. Rev. Lett.}\ }\textbf {\bibinfo {volume} {90}},\ \bibinfo {pages}
  {100401} (\bibinfo {year} {2003})}\BibitemShut {NoStop}%
\bibitem [{\citenamefont {Kuklov}\ \emph {et~al.}(2004)\citenamefont {Kuklov},
  \citenamefont {Prokof'ev},\ and\ \citenamefont {Svistunov}}]{Kuklov2004}%
  \BibitemOpen
  \bibfield  {author} {\bibinfo {author} {\bibfnamefont {A.}~\bibnamefont
  {Kuklov}}, \bibinfo {author} {\bibfnamefont {N.}~\bibnamefont {Prokof'ev}}, \
  and\ \bibinfo {author} {\bibfnamefont {B.}~\bibnamefont {Svistunov}},\ }\href
  {\doibase 10.1103/PhysRevLett.92.050402} {\bibfield  {journal} {\bibinfo
  {journal} {Phys. Rev. Lett.}\ }\textbf {\bibinfo {volume} {92}},\ \bibinfo
  {pages} {050402} (\bibinfo {year} {2004})}\BibitemShut {NoStop}%
\bibitem [{\citenamefont {Dahl}\ \emph {et~al.}(2008)\citenamefont {Dahl},
  \citenamefont {Babaev}, \citenamefont {Kragset},\ and\ \citenamefont
  {Sudb\"o}}]{Dahl2008_preemptive}%
  \BibitemOpen
  \bibfield  {author} {\bibinfo {author} {\bibfnamefont {E.~K.}\ \bibnamefont
  {Dahl}}, \bibinfo {author} {\bibfnamefont {E.}~\bibnamefont {Babaev}},
  \bibinfo {author} {\bibfnamefont {S.}~\bibnamefont {Kragset}}, \ and\
  \bibinfo {author} {\bibfnamefont {A.}~\bibnamefont {Sudb\"o}},\ }\href
  {\doibase 10.1103/PhysRevB.77.144519} {\bibfield  {journal} {\bibinfo
  {journal} {Phys. Rev. B}\ }\textbf {\bibinfo {volume} {77}},\ \bibinfo
  {pages} {144519} (\bibinfo {year} {2008})}\BibitemShut {NoStop}%
\bibitem [{\citenamefont {Kuklov}\ \emph {et~al.}(2006)\citenamefont {Kuklov},
  \citenamefont {Prokof'ev}, \citenamefont {Svistunov},\ and\ \citenamefont
  {Troyer}}]{Kuklov2006}%
  \BibitemOpen
  \bibfield  {author} {\bibinfo {author} {\bibfnamefont {A.}~\bibnamefont
  {Kuklov}}, \bibinfo {author} {\bibfnamefont {N.}~\bibnamefont {Prokof'ev}},
  \bibinfo {author} {\bibfnamefont {B.}~\bibnamefont {Svistunov}}, \ and\
  \bibinfo {author} {\bibfnamefont {M.}~\bibnamefont {Troyer}},\ }\href
  {\doibase https://doi.org/10.1016/j.aop.2006.04.007} {\bibfield  {journal}
  {\bibinfo  {journal} {Ann. Phys.}\ }\textbf {\bibinfo {volume} {321}},\
  \bibinfo {pages} {1602 } (\bibinfo {year} {2006})},\ \bibinfo {note} {july
  2006 Special Issue}\BibitemShut {NoStop}%
\bibitem [{\citenamefont {Kuklov}\ \emph {et~al.}(2008)\citenamefont {Kuklov},
  \citenamefont {Matsumoto}, \citenamefont {Prokof'ev}, \citenamefont
  {Svistunov},\ and\ \citenamefont {Troyer}}]{Kuklov2008_deconfined}%
  \BibitemOpen
  \bibfield  {author} {\bibinfo {author} {\bibfnamefont {A.~B.}\ \bibnamefont
  {Kuklov}}, \bibinfo {author} {\bibfnamefont {M.}~\bibnamefont {Matsumoto}},
  \bibinfo {author} {\bibfnamefont {N.~V.}\ \bibnamefont {Prokof'ev}}, \bibinfo
  {author} {\bibfnamefont {B.~V.}\ \bibnamefont {Svistunov}}, \ and\ \bibinfo
  {author} {\bibfnamefont {M.}~\bibnamefont {Troyer}},\ }\href {\doibase
  10.1103/PhysRevLett.101.050405} {\bibfield  {journal} {\bibinfo  {journal}
  {Phys. Rev. Lett.}\ }\textbf {\bibinfo {volume} {101}},\ \bibinfo {pages}
  {050405} (\bibinfo {year} {2008})}\BibitemShut {NoStop}%
\bibitem [{\citenamefont {Grinenko}\ \emph
  {et~al.}(2021{\natexlab{a}})\citenamefont {Grinenko}, \citenamefont {Weston},
  \citenamefont {Caglieris}, \citenamefont {Wuttke}, \citenamefont {Hess},
  \citenamefont {Gottschall}, \citenamefont {Maccari}, \citenamefont
  {Gorbunov}, \citenamefont {Zherlitsyn}, \citenamefont {Wosnitza},
  \citenamefont {Rydh}, \citenamefont {Kihou}, \citenamefont {Lee},
  \citenamefont {Sarkar}, \citenamefont {Dengre}, \citenamefont {Garaud},
  \citenamefont {Charnukha}, \citenamefont {Hühne}, \citenamefont {Nielsch},
  \citenamefont {Büchner}, \citenamefont {Klauss},\ and\ \citenamefont
  {Babaev}}]{Grinenko2021_state}%
  \BibitemOpen
  \bibfield  {author} {\bibinfo {author} {\bibfnamefont {V.}~\bibnamefont
  {Grinenko}}, \bibinfo {author} {\bibfnamefont {D.}~\bibnamefont {Weston}},
  \bibinfo {author} {\bibfnamefont {F.}~\bibnamefont {Caglieris}}, \bibinfo
  {author} {\bibfnamefont {C.}~\bibnamefont {Wuttke}}, \bibinfo {author}
  {\bibfnamefont {C.}~\bibnamefont {Hess}}, \bibinfo {author} {\bibfnamefont
  {T.}~\bibnamefont {Gottschall}}, \bibinfo {author} {\bibfnamefont
  {I.}~\bibnamefont {Maccari}}, \bibinfo {author} {\bibfnamefont
  {D.}~\bibnamefont {Gorbunov}}, \bibinfo {author} {\bibfnamefont
  {S.}~\bibnamefont {Zherlitsyn}}, \bibinfo {author} {\bibfnamefont
  {J.}~\bibnamefont {Wosnitza}}, \bibinfo {author} {\bibfnamefont
  {A.}~\bibnamefont {Rydh}}, \bibinfo {author} {\bibfnamefont {K.}~\bibnamefont
  {Kihou}}, \bibinfo {author} {\bibfnamefont {C.-H.}\ \bibnamefont {Lee}},
  \bibinfo {author} {\bibfnamefont {R.}~\bibnamefont {Sarkar}}, \bibinfo
  {author} {\bibfnamefont {S.}~\bibnamefont {Dengre}}, \bibinfo {author}
  {\bibfnamefont {J.}~\bibnamefont {Garaud}}, \bibinfo {author} {\bibfnamefont
  {A.}~\bibnamefont {Charnukha}}, \bibinfo {author} {\bibfnamefont
  {R.}~\bibnamefont {Hühne}}, \bibinfo {author} {\bibfnamefont
  {K.}~\bibnamefont {Nielsch}}, \bibinfo {author} {\bibfnamefont
  {B.}~\bibnamefont {Büchner}}, \bibinfo {author} {\bibfnamefont {H.-H.}\
  \bibnamefont {Klauss}}, \ and\ \bibinfo {author} {\bibfnamefont
  {E.}~\bibnamefont {Babaev}},\ }\href {\doibase 10.1038/s41567-021-01350-9}
  {\bibfield  {journal} {\bibinfo  {journal} {Nature Physics}\ ,\ \bibinfo
  {pages} {1}} (\bibinfo {year} {2021}{\natexlab{a}})}\BibitemShut {NoStop}%
\bibitem [{\citenamefont {Ng}\ and\ \citenamefont {Nagaosa}(2009)}]{Ng2009}%
  \BibitemOpen
  \bibfield  {author} {\bibinfo {author} {\bibfnamefont {T.~K.}\ \bibnamefont
  {Ng}}\ and\ \bibinfo {author} {\bibfnamefont {N.}~\bibnamefont {Nagaosa}},\
  }\href {\doibase 10.1209/0295-5075/87/17003} {\bibfield  {journal} {\bibinfo
  {journal} {EPL}\ }\textbf {\bibinfo {volume} {87}},\ \bibinfo {pages} {17003}
  (\bibinfo {year} {2009})}\BibitemShut {NoStop}%
\bibitem [{\citenamefont {Stanev}\ and\ \citenamefont
  {Tesanovic}(2010)}]{Stanev2010}%
  \BibitemOpen
  \bibfield  {author} {\bibinfo {author} {\bibfnamefont {V.}~\bibnamefont
  {Stanev}}\ and\ \bibinfo {author} {\bibfnamefont {Z.}~\bibnamefont
  {Tesanovic}},\ }\href {\doibase 10.1103/PhysRevB.81.134522} {\bibfield
  {journal} {\bibinfo  {journal} {Phys. Rev. B}\ }\textbf {\bibinfo {volume}
  {81}},\ \bibinfo {pages} {134522} (\bibinfo {year} {2010})}\BibitemShut
  {NoStop}%
\bibitem [{\citenamefont {Carlstr{\"o}m}\ \emph {et~al.}(2011)\citenamefont
  {Carlstr{\"o}m}, \citenamefont {Garaud},\ and\ \citenamefont
  {Babaev}}]{Carlstrom2011_lengthscales}%
  \BibitemOpen
  \bibfield  {author} {\bibinfo {author} {\bibfnamefont {J.}~\bibnamefont
  {Carlstr{\"o}m}}, \bibinfo {author} {\bibfnamefont {J.}~\bibnamefont
  {Garaud}}, \ and\ \bibinfo {author} {\bibfnamefont {E.}~\bibnamefont
  {Babaev}},\ }\href {\doibase 10.1103/PhysRevB.84.134518} {\bibfield
  {journal} {\bibinfo  {journal} {Phys. Rev. B}\ }\textbf {\bibinfo {volume}
  {84}},\ \bibinfo {pages} {134518} (\bibinfo {year} {2011})}\BibitemShut
  {NoStop}%
\bibitem [{\citenamefont {{Maiti}}\ and\ \citenamefont
  {{Chubukov}}(2013)}]{Maiti2013}%
  \BibitemOpen
  \bibfield  {author} {\bibinfo {author} {\bibfnamefont {S.}~\bibnamefont
  {{Maiti}}}\ and\ \bibinfo {author} {\bibfnamefont {A.~V.}\ \bibnamefont
  {{Chubukov}}},\ }\href {\doibase 10.1103/PhysRevB.87.144511} {\bibfield
  {journal} {\bibinfo  {journal} {Phys. Rev. B}\ }\textbf {\bibinfo {volume}
  {87}},\ \bibinfo {eid} {144511} (\bibinfo {year} {2013})}\BibitemShut
  {NoStop}%
\bibitem [{\citenamefont {Silaev}\ \emph {et~al.}(2017)\citenamefont {Silaev},
  \citenamefont {Garaud},\ and\ \citenamefont {Babaev}}]{silaev2017phase}%
  \BibitemOpen
  \bibfield  {author} {\bibinfo {author} {\bibfnamefont {M.}~\bibnamefont
  {Silaev}}, \bibinfo {author} {\bibfnamefont {J.}~\bibnamefont {Garaud}}, \
  and\ \bibinfo {author} {\bibfnamefont {E.}~\bibnamefont {Babaev}},\
  }\href@noop {} {\bibfield  {journal} {\bibinfo  {journal} {Phys. Rev. B}\
  }\textbf {\bibinfo {volume} {95}},\ \bibinfo {pages} {024517} (\bibinfo
  {year} {2017})}\BibitemShut {NoStop}%
\bibitem [{\citenamefont {B\"oker}\ \emph {et~al.}(2017)\citenamefont
  {B\"oker}, \citenamefont {Volkov}, \citenamefont {Efetov},\ and\
  \citenamefont {Eremin}}]{Boeker2017}%
  \BibitemOpen
  \bibfield  {author} {\bibinfo {author} {\bibfnamefont {J.}~\bibnamefont
  {B\"oker}}, \bibinfo {author} {\bibfnamefont {P.~A.}\ \bibnamefont {Volkov}},
  \bibinfo {author} {\bibfnamefont {K.~B.}\ \bibnamefont {Efetov}}, \ and\
  \bibinfo {author} {\bibfnamefont {I.}~\bibnamefont {Eremin}},\ }\href
  {\doibase 10.1103/PhysRevB.96.014517} {\bibfield  {journal} {\bibinfo
  {journal} {Phys. Rev. B}\ }\textbf {\bibinfo {volume} {96}},\ \bibinfo
  {pages} {014517} (\bibinfo {year} {2017})}\BibitemShut {NoStop}%
\bibitem [{\citenamefont {Grinenko}\ \emph {et~al.}(2017)\citenamefont
  {Grinenko}, \citenamefont {Materne}, \citenamefont {Sarkar}, \citenamefont
  {Luetkens}, \citenamefont {Kihou}, \citenamefont {Lee}, \citenamefont
  {Akhmadaliev}, \citenamefont {Efremov}, \citenamefont {Drechsler},\ and\
  \citenamefont {Klauss}}]{Grinenko2017}%
  \BibitemOpen
  \bibfield  {author} {\bibinfo {author} {\bibfnamefont {V.}~\bibnamefont
  {Grinenko}}, \bibinfo {author} {\bibfnamefont {P.}~\bibnamefont {Materne}},
  \bibinfo {author} {\bibfnamefont {R.}~\bibnamefont {Sarkar}}, \bibinfo
  {author} {\bibfnamefont {H.}~\bibnamefont {Luetkens}}, \bibinfo {author}
  {\bibfnamefont {K.}~\bibnamefont {Kihou}}, \bibinfo {author} {\bibfnamefont
  {C.~H.}\ \bibnamefont {Lee}}, \bibinfo {author} {\bibfnamefont
  {S.}~\bibnamefont {Akhmadaliev}}, \bibinfo {author} {\bibfnamefont {D.~V.}\
  \bibnamefont {Efremov}}, \bibinfo {author} {\bibfnamefont {S.-L.}\
  \bibnamefont {Drechsler}}, \ and\ \bibinfo {author} {\bibfnamefont {H.-H.}\
  \bibnamefont {Klauss}},\ }\href {\doibase 10.1103/PhysRevB.95.214511}
  {\bibfield  {journal} {\bibinfo  {journal} {Phys. Rev. B}\ }\textbf {\bibinfo
  {volume} {95}},\ \bibinfo {pages} {214511} (\bibinfo {year}
  {2017})}\BibitemShut {NoStop}%
\bibitem [{\citenamefont {Kivelson}\ \emph {et~al.}(2020)\citenamefont
  {Kivelson}, \citenamefont {Yuan}, \citenamefont {Ramshaw},\ and\
  \citenamefont {Thomale}}]{Kivelson2020}%
  \BibitemOpen
  \bibfield  {author} {\bibinfo {author} {\bibfnamefont {S.}~\bibnamefont
  {Kivelson}}, \bibinfo {author} {\bibfnamefont {A.}~\bibnamefont {Yuan}},
  \bibinfo {author} {\bibfnamefont {B.}~\bibnamefont {Ramshaw}}, \ and\
  \bibinfo {author} {\bibfnamefont {R.}~\bibnamefont {Thomale}},\ }\href@noop
  {} {\bibfield  {journal} {\bibinfo  {journal} {npj Quantum Mat.}\ }\textbf
  {\bibinfo {volume} {5}},\ \bibinfo {pages} {43} (\bibinfo {year}
  {2020})}\BibitemShut {NoStop}%
\bibitem [{\citenamefont {Grinenko}\ \emph
  {et~al.}(2021{\natexlab{b}})\citenamefont {Grinenko}, \citenamefont {Ghosh},
  \citenamefont {Sarkar}, \citenamefont {Orain}, \citenamefont {Nikitin},
  \citenamefont {Elender}, \citenamefont {Das}, \citenamefont {Guguchia},
  \citenamefont {Bruckner}, \citenamefont {Barber}, \citenamefont {Park},
  \citenamefont {Kikugawa}, \citenamefont {Sokolov}, \citenamefont {Bobowski},
  \citenamefont {Miyoshi}, \citenamefont {Maeno}, \citenamefont {Mackenzie},
  \citenamefont {Luetkens}, \citenamefont {Hicks},\ and\ \citenamefont
  {Klauss}}]{Grinenko2020}%
  \BibitemOpen
  \bibfield  {author} {\bibinfo {author} {\bibfnamefont {V.}~\bibnamefont
  {Grinenko}}, \bibinfo {author} {\bibfnamefont {S.}~\bibnamefont {Ghosh}},
  \bibinfo {author} {\bibfnamefont {R.}~\bibnamefont {Sarkar}}, \bibinfo
  {author} {\bibfnamefont {J.-C.}\ \bibnamefont {Orain}}, \bibinfo {author}
  {\bibfnamefont {A.}~\bibnamefont {Nikitin}}, \bibinfo {author} {\bibfnamefont
  {M.}~\bibnamefont {Elender}}, \bibinfo {author} {\bibfnamefont
  {D.}~\bibnamefont {Das}}, \bibinfo {author} {\bibfnamefont {Z.}~\bibnamefont
  {Guguchia}}, \bibinfo {author} {\bibfnamefont {F.}~\bibnamefont {Bruckner}},
  \bibinfo {author} {\bibfnamefont {M.~E.}\ \bibnamefont {Barber}}, \bibinfo
  {author} {\bibfnamefont {J.}~\bibnamefont {Park}}, \bibinfo {author}
  {\bibfnamefont {N.}~\bibnamefont {Kikugawa}}, \bibinfo {author}
  {\bibfnamefont {D.~A.}\ \bibnamefont {Sokolov}}, \bibinfo {author}
  {\bibfnamefont {J.~S.}\ \bibnamefont {Bobowski}}, \bibinfo {author}
  {\bibfnamefont {T.}~\bibnamefont {Miyoshi}}, \bibinfo {author} {\bibfnamefont
  {Y.}~\bibnamefont {Maeno}}, \bibinfo {author} {\bibfnamefont {A.~P.}\
  \bibnamefont {Mackenzie}}, \bibinfo {author} {\bibfnamefont {H.}~\bibnamefont
  {Luetkens}}, \bibinfo {author} {\bibfnamefont {C.~W.}\ \bibnamefont {Hicks}},
  \ and\ \bibinfo {author} {\bibfnamefont {H.-H.}\ \bibnamefont {Klauss}},\
  }\href {\doibase 10.1038/s41567-021-01182-7} {\bibfield  {journal} {\bibinfo
  {journal} {Nat. Phys.}\ } (\bibinfo {year} {2021}{\natexlab{b}}),\
  10.1038/s41567-021-01182-7}\BibitemShut {NoStop}%
\bibitem [{\citenamefont {Grinenko}\ \emph
  {et~al.}(2021{\natexlab{c}})\citenamefont {Grinenko}, \citenamefont {Das},
  \citenamefont {Gupta}, \citenamefont {Zinkl}, \citenamefont {Kikugawa},
  \citenamefont {Maeno}, \citenamefont {Hicks}, \citenamefont {Klauss},
  \citenamefont {Sigrist},\ and\ \citenamefont
  {Khasanov}}]{Grinenko2021_unsplit}%
  \BibitemOpen
  \bibfield  {author} {\bibinfo {author} {\bibfnamefont {V.}~\bibnamefont
  {Grinenko}}, \bibinfo {author} {\bibfnamefont {D.}~\bibnamefont {Das}},
  \bibinfo {author} {\bibfnamefont {R.}~\bibnamefont {Gupta}}, \bibinfo
  {author} {\bibfnamefont {B.}~\bibnamefont {Zinkl}}, \bibinfo {author}
  {\bibfnamefont {N.}~\bibnamefont {Kikugawa}}, \bibinfo {author}
  {\bibfnamefont {Y.}~\bibnamefont {Maeno}}, \bibinfo {author} {\bibfnamefont
  {C.~W.}\ \bibnamefont {Hicks}}, \bibinfo {author} {\bibfnamefont {H.-H.}\
  \bibnamefont {Klauss}}, \bibinfo {author} {\bibfnamefont {M.}~\bibnamefont
  {Sigrist}}, \ and\ \bibinfo {author} {\bibfnamefont {R.}~\bibnamefont
  {Khasanov}},\ }\href {https://www.nature.com/articles/s41467-021-24176-8}
  {\bibfield  {journal} {\bibinfo  {journal} {Nature Communications}\ }\textbf
  {\bibinfo {volume} {12}},\ \bibinfo {pages} {3920} (\bibinfo {year}
  {2021}{\natexlab{c}})}\BibitemShut {NoStop}%
\bibitem [{\citenamefont {Grinenko}\ \emph {et~al.}(2020)\citenamefont
  {Grinenko}, \citenamefont {Sarkar}, \citenamefont {Kihou}, \citenamefont
  {Lee}, \citenamefont {Morozov}, \citenamefont {Aswartham}, \citenamefont
  {B{\"u}chner}, \citenamefont {Chekhonin}, \citenamefont {Skrotzki},
  \citenamefont {Nenkov}, \citenamefont {H{\"u}hne}, \citenamefont {Nielsch},
  \citenamefont {Drechsler}, \citenamefont {Vadimov}, \citenamefont {Silaev},
  \citenamefont {Volkov}, \citenamefont {Eremin}, \citenamefont {Luetkens},\
  and\ \citenamefont {Klauss}}]{Grinenko2018}%
  \BibitemOpen
  \bibfield  {author} {\bibinfo {author} {\bibfnamefont {V.}~\bibnamefont
  {Grinenko}}, \bibinfo {author} {\bibfnamefont {R.}~\bibnamefont {Sarkar}},
  \bibinfo {author} {\bibfnamefont {K.}~\bibnamefont {Kihou}}, \bibinfo
  {author} {\bibfnamefont {C.~H.}\ \bibnamefont {Lee}}, \bibinfo {author}
  {\bibfnamefont {I.}~\bibnamefont {Morozov}}, \bibinfo {author} {\bibfnamefont
  {S.}~\bibnamefont {Aswartham}}, \bibinfo {author} {\bibfnamefont
  {B.}~\bibnamefont {B{\"u}chner}}, \bibinfo {author} {\bibfnamefont
  {P.}~\bibnamefont {Chekhonin}}, \bibinfo {author} {\bibfnamefont
  {W.}~\bibnamefont {Skrotzki}}, \bibinfo {author} {\bibfnamefont
  {K.}~\bibnamefont {Nenkov}}, \bibinfo {author} {\bibfnamefont
  {R.}~\bibnamefont {H{\"u}hne}}, \bibinfo {author} {\bibfnamefont
  {K.}~\bibnamefont {Nielsch}}, \bibinfo {author} {\bibfnamefont {S.~L.}\
  \bibnamefont {Drechsler}}, \bibinfo {author} {\bibfnamefont {V.~L.}\
  \bibnamefont {Vadimov}}, \bibinfo {author} {\bibfnamefont {M.~A.}\
  \bibnamefont {Silaev}}, \bibinfo {author} {\bibfnamefont {P.}~\bibnamefont
  {Volkov}}, \bibinfo {author} {\bibfnamefont {I.}~\bibnamefont {Eremin}},
  \bibinfo {author} {\bibfnamefont {H.}~\bibnamefont {Luetkens}}, \ and\
  \bibinfo {author} {\bibfnamefont {H.~H.}\ \bibnamefont {Klauss}},\
  }\href@noop {} {\bibfield  {journal} {\bibinfo  {journal} {Nat. Phys.}\
  }\textbf {\bibinfo {volume} {16}},\ \bibinfo {pages} {789–794} (\bibinfo
  {year} {2020})}\BibitemShut {NoStop}%
\bibitem [{\citenamefont {Vadimov}\ and\ \citenamefont
  {Silaev}(2018)}]{Vadimov2018}%
  \BibitemOpen
  \bibfield  {author} {\bibinfo {author} {\bibfnamefont {V.~L.}\ \bibnamefont
  {Vadimov}}\ and\ \bibinfo {author} {\bibfnamefont {M.~A.}\ \bibnamefont
  {Silaev}},\ }\href {\doibase 10.1103/PhysRevB.98.104504} {\bibfield
  {journal} {\bibinfo  {journal} {Phys. Rev. B}\ }\textbf {\bibinfo {volume}
  {98}},\ \bibinfo {pages} {104504} (\bibinfo {year} {2018})}\BibitemShut
  {NoStop}%
\bibitem [{\citenamefont {Garaud}\ and\ \citenamefont
  {Babaev}(2014)}]{garaud2014domain}%
  \BibitemOpen
  \bibfield  {author} {\bibinfo {author} {\bibfnamefont {J.}~\bibnamefont
  {Garaud}}\ and\ \bibinfo {author} {\bibfnamefont {E.}~\bibnamefont
  {Babaev}},\ }\href@noop {} {\bibfield  {journal} {\bibinfo  {journal} {Phys.
  Rev. Lett.}\ }\textbf {\bibinfo {volume} {112}},\ \bibinfo {pages} {017003}
  (\bibinfo {year} {2014})}\BibitemShut {NoStop}%
\bibitem [{\citenamefont {Lee}\ \emph {et~al.}(2009)\citenamefont {Lee},
  \citenamefont {Zhang},\ and\ \citenamefont {Wu}}]{lee2009pairing}%
  \BibitemOpen
  \bibfield  {author} {\bibinfo {author} {\bibfnamefont {W.-C.}\ \bibnamefont
  {Lee}}, \bibinfo {author} {\bibfnamefont {S.-C.}\ \bibnamefont {Zhang}}, \
  and\ \bibinfo {author} {\bibfnamefont {C.}~\bibnamefont {Wu}},\ }\href@noop
  {} {\bibfield  {journal} {\bibinfo  {journal} {Physical review letters}\
  }\textbf {\bibinfo {volume} {102}},\ \bibinfo {pages} {217002} (\bibinfo
  {year} {2009})}\BibitemShut {NoStop}%
\bibitem [{\citenamefont {Garaud}\ and\ \citenamefont
  {Babaev}(2021)}]{garaud2021skyrmions}%
  \BibitemOpen
  \bibfield  {author} {\bibinfo {author} {\bibfnamefont {J.}~\bibnamefont
  {Garaud}}\ and\ \bibinfo {author} {\bibfnamefont {E.}~\bibnamefont
  {Babaev}},\ }\href@noop {} {\bibfield  {journal} {\bibinfo  {journal} {arXiv
  preprint arXiv:2112.01286}\ } (\bibinfo {year} {2021})}\BibitemShut {NoStop}%
\bibitem [{\citenamefont {Garaud}\ \emph {et~al.}(2017)\citenamefont {Garaud},
  \citenamefont {Silaev},\ and\ \citenamefont {Babaev}}]{Garaud2017}%
  \BibitemOpen
  \bibfield  {author} {\bibinfo {author} {\bibfnamefont {J.}~\bibnamefont
  {Garaud}}, \bibinfo {author} {\bibfnamefont {M.}~\bibnamefont {Silaev}}, \
  and\ \bibinfo {author} {\bibfnamefont {E.}~\bibnamefont {Babaev}},\ }\href
  {\doibase https://doi.org/10.1016/j.physc.2016.07.010} {\bibfield  {journal}
  {\bibinfo  {journal} {Physica C}\ }\textbf {\bibinfo {volume} {533}},\
  \bibinfo {pages} {63 } (\bibinfo {year} {2017})},\ \bibinfo {note} {ninth
  international conference on Vortex Matter in nanostructured
  Superdonductors}\BibitemShut {NoStop}%
\bibitem [{\citenamefont {Garaud}\ \emph {et~al.}(2018)\citenamefont {Garaud},
  \citenamefont {Corticelli}, \citenamefont {Silaev},\ and\ \citenamefont
  {Babaev}}]{garaud2018properties}%
  \BibitemOpen
  \bibfield  {author} {\bibinfo {author} {\bibfnamefont {J.}~\bibnamefont
  {Garaud}}, \bibinfo {author} {\bibfnamefont {A.}~\bibnamefont {Corticelli}},
  \bibinfo {author} {\bibfnamefont {M.}~\bibnamefont {Silaev}}, \ and\ \bibinfo
  {author} {\bibfnamefont {E.}~\bibnamefont {Babaev}},\ }\href@noop {}
  {\bibfield  {journal} {\bibinfo  {journal} {Phys. Rev. B}\ }\textbf {\bibinfo
  {volume} {98}},\ \bibinfo {pages} {014520} (\bibinfo {year}
  {2018})}\BibitemShut {NoStop}%
\bibitem [{\citenamefont {Agterberg}(1998)}]{Agterberg1998vortex}%
  \BibitemOpen
  \bibfield  {author} {\bibinfo {author} {\bibfnamefont {D.~F.}\ \bibnamefont
  {Agterberg}},\ }\href {\doibase 10.1103/PhysRevLett.80.5184} {\bibfield
  {journal} {\bibinfo  {journal} {Phys. Rev. Lett.}\ }\textbf {\bibinfo
  {volume} {80}},\ \bibinfo {pages} {5184} (\bibinfo {year}
  {1998})}\BibitemShut {NoStop}%
\bibitem [{\citenamefont {Haugen}\ \emph {et~al.}(2021)\citenamefont {Haugen},
  \citenamefont {Babaev}, \citenamefont {Krohg},\ and\ \citenamefont
  {Sudb\"o}}]{Haugen2021first}%
  \BibitemOpen
  \bibfield  {author} {\bibinfo {author} {\bibfnamefont {H.~H.}\ \bibnamefont
  {Haugen}}, \bibinfo {author} {\bibfnamefont {E.}~\bibnamefont {Babaev}},
  \bibinfo {author} {\bibfnamefont {F.~N.}\ \bibnamefont {Krohg}}, \ and\
  \bibinfo {author} {\bibfnamefont {A.}~\bibnamefont {Sudb\"o}},\ }\href
  {\doibase 10.1103/PhysRevB.104.104515} {\bibfield  {journal} {\bibinfo
  {journal} {Phys. Rev. B}\ }\textbf {\bibinfo {volume} {104}},\ \bibinfo
  {pages} {104515} (\bibinfo {year} {2021})}\BibitemShut {NoStop}%
\bibitem [{\citenamefont {Leggett}(1975)}]{leggett1975theoretical}%
  \BibitemOpen
  \bibfield  {author} {\bibinfo {author} {\bibfnamefont {A.~J.}\ \bibnamefont
  {Leggett}},\ }\href@noop {} {\bibfield  {journal} {\bibinfo  {journal} {Rev.
  Mod. Phys.}\ }\textbf {\bibinfo {volume} {47}},\ \bibinfo {pages} {331}
  (\bibinfo {year} {1975})}\BibitemShut {NoStop}%
\bibitem [{\citenamefont {Sj\"oberg}(1976)}]{Sjoberg:76}%
  \BibitemOpen
  \bibfield  {author} {\bibinfo {author} {\bibfnamefont {O.}~\bibnamefont
  {Sj\"oberg}},\ }\href {\doibase 10.1016/0375-9474(76)90558-3} {\bibfield
  {journal} {\bibinfo  {journal} {Nuclear Physics A}\ }\textbf {\bibinfo
  {volume} {265}},\ \bibinfo {pages} {511} (\bibinfo {year}
  {1976})}\BibitemShut {NoStop}%
\bibitem [{\citenamefont {Svistunov}\ \emph {et~al.}(2015)\citenamefont
  {Svistunov}, \citenamefont {Babaev},\ and\ \citenamefont
  {Prokofev}}]{Svistunov2015}%
  \BibitemOpen
  \bibfield  {author} {\bibinfo {author} {\bibfnamefont {B.}~\bibnamefont
  {Svistunov}}, \bibinfo {author} {\bibfnamefont {E.}~\bibnamefont {Babaev}}, \
  and\ \bibinfo {author} {\bibfnamefont {N.}~\bibnamefont {Prokofev}},\
  }\href@noop {} {\emph {\bibinfo {title} {Superfluid States of Matter}}}\
  (\bibinfo  {publisher} {CRC Press},\ \bibinfo {year} {2015})\BibitemShut
  {NoStop}%
\bibitem [{\citenamefont {Sellin}\ and\ \citenamefont
  {Babaev}(2018)}]{Sellin.Babaev:18}%
  \BibitemOpen
  \bibfield  {author} {\bibinfo {author} {\bibfnamefont {K.}~\bibnamefont
  {Sellin}}\ and\ \bibinfo {author} {\bibfnamefont {E.}~\bibnamefont
  {Babaev}},\ }\href {\doibase 10.1103/PhysRevB.97.094517} {\bibfield
  {journal} {\bibinfo  {journal} {Phys. Rev. B}\ }\textbf {\bibinfo {volume}
  {97}},\ \bibinfo {pages} {094517} (\bibinfo {year} {2018})}\BibitemShut
  {NoStop}%
\bibitem [{\citenamefont {Hartman}\ \emph {et~al.}(2018)\citenamefont
  {Hartman}, \citenamefont {Erlandsen},\ and\ \citenamefont
  {Sudb\"o}}]{hartman2018superfluid}%
  \BibitemOpen
  \bibfield  {author} {\bibinfo {author} {\bibfnamefont {S.}~\bibnamefont
  {Hartman}}, \bibinfo {author} {\bibfnamefont {E.}~\bibnamefont {Erlandsen}},
  \ and\ \bibinfo {author} {\bibfnamefont {A.}~\bibnamefont {Sudb\"o}},\ }\href
  {\doibase 10.1103/PhysRevB.98.024512} {\bibfield  {journal} {\bibinfo
  {journal} {Phys. Rev. B}\ }\textbf {\bibinfo {volume} {98}},\ \bibinfo
  {pages} {024512} (\bibinfo {year} {2018})}\BibitemShut {NoStop}%
\bibitem [{\citenamefont {Linder}\ and\ \citenamefont
  {Sudb\"o}(2009)}]{linder2009calculation}%
  \BibitemOpen
  \bibfield  {author} {\bibinfo {author} {\bibfnamefont {J.}~\bibnamefont
  {Linder}}\ and\ \bibinfo {author} {\bibfnamefont {A.}~\bibnamefont
  {Sudb\"o}},\ }\href {\doibase 10.1103/PhysRevA.79.063610} {\bibfield
  {journal} {\bibinfo  {journal} {Phys. Rev. A}\ }\textbf {\bibinfo {volume}
  {79}},\ \bibinfo {pages} {063610} (\bibinfo {year} {2009})}\BibitemShut
  {NoStop}%
\bibitem [{\citenamefont {{Smiseth}}\ \emph {et~al.}(2005)\citenamefont
  {{Smiseth}}, \citenamefont {{Sm\"orgrav}}, \citenamefont {{Babaev}},\ and\
  \citenamefont {{Sudb\"o}}}]{Smiseth2005}%
  \BibitemOpen
  \bibfield  {author} {\bibinfo {author} {\bibfnamefont {J.}~\bibnamefont
  {{Smiseth}}}, \bibinfo {author} {\bibfnamefont {E.}~\bibnamefont
  {{Sm\"orgrav}}}, \bibinfo {author} {\bibfnamefont {E.}~\bibnamefont
  {{Babaev}}}, \ and\ \bibinfo {author} {\bibfnamefont {A.}~\bibnamefont
  {{Sudb\"o}}},\ }\href {\doibase 10.1103/PhysRevB.71.214509} {\bibfield
  {journal} {\bibinfo  {journal} {Phys. Rev. B}\ }\textbf {\bibinfo {volume}
  {71}},\ \bibinfo {eid} {214509} (\bibinfo {year} {2005})}\BibitemShut
  {NoStop}%
\bibitem [{\citenamefont {Kuklov}\ \emph {et~al.}(2005)\citenamefont {Kuklov},
  \citenamefont {Prokof'ev},\ and\ \citenamefont {Svistunov}}]{Kuklov2005}%
  \BibitemOpen
  \bibfield  {author} {\bibinfo {author} {\bibfnamefont {A.}~\bibnamefont
  {Kuklov}}, \bibinfo {author} {\bibfnamefont {N.}~\bibnamefont {Prokof'ev}}, \
  and\ \bibinfo {author} {\bibfnamefont {B.}~\bibnamefont {Svistunov}},\
  }\href@noop {} {\bibfield  {journal} {\bibinfo  {journal} {arXiv}\ }
  (\bibinfo {year} {2005})},\ \Eprint {http://arxiv.org/abs/cond-mat/0501052v2}
  {cond-mat/0501052v2} \BibitemShut {NoStop}%
\bibitem [{\citenamefont {Blomquist}\ \emph {et~al.}(2021)\citenamefont
  {Blomquist}, \citenamefont {Syrwid},\ and\ \citenamefont
  {Babaev}}]{blomquist2021borromean}%
  \BibitemOpen
  \bibfield  {author} {\bibinfo {author} {\bibfnamefont {E.}~\bibnamefont
  {Blomquist}}, \bibinfo {author} {\bibfnamefont {A.}~\bibnamefont {Syrwid}}, \
  and\ \bibinfo {author} {\bibfnamefont {E.}~\bibnamefont {Babaev}},\
  }\href@noop {} {\bibfield  {journal} {\bibinfo  {journal} {Physical review
  letters}\ }\textbf {\bibinfo {volume} {127}},\ \bibinfo {pages} {255303}
  (\bibinfo {year} {2021})}\BibitemShut {NoStop}%
\bibitem [{\citenamefont {Babaev}\ \emph {et~al.}(2002)\citenamefont {Babaev},
  \citenamefont {Faddeev},\ and\ \citenamefont {Niemi}}]{Babaev2002_hidden}%
  \BibitemOpen
  \bibfield  {author} {\bibinfo {author} {\bibfnamefont {E.}~\bibnamefont
  {Babaev}}, \bibinfo {author} {\bibfnamefont {L.~D.}\ \bibnamefont {Faddeev}},
  \ and\ \bibinfo {author} {\bibfnamefont {A.~J.}\ \bibnamefont {Niemi}},\
  }\href {\doibase 10.1103/PhysRevB.65.100512 (R)} {\bibfield  {journal}
  {\bibinfo  {journal} {Phys. Rev. B}\ }\textbf {\bibinfo {volume} {65}},\
  \bibinfo {pages} {100512} (\bibinfo {year} {2002})}\BibitemShut {NoStop}%
\bibitem [{\citenamefont {Babaev}(2002{\natexlab{b}})}]{Babaev2002_vortices}%
  \BibitemOpen
  \bibfield  {author} {\bibinfo {author} {\bibfnamefont {E.}~\bibnamefont
  {Babaev}},\ }\href {\doibase 10.1103/PhysRevLett.89.067001} {\bibfield
  {journal} {\bibinfo  {journal} {Phys. Rev. Lett.}\ }\textbf {\bibinfo
  {volume} {89}},\ \bibinfo {pages} {067001} (\bibinfo {year}
  {2002}{\natexlab{b}})}\BibitemShut {NoStop}%
\bibitem [{\citenamefont {{Garaud}}\ \emph {et~al.}(2011)\citenamefont
  {{Garaud}}, \citenamefont {{Carlstr{\"o}m}},\ and\ \citenamefont
  {{Babaev}}}]{Garaud2011topological}%
  \BibitemOpen
  \bibfield  {author} {\bibinfo {author} {\bibfnamefont {J.}~\bibnamefont
  {{Garaud}}}, \bibinfo {author} {\bibfnamefont {J.}~\bibnamefont
  {{Carlstr{\"o}m}}}, \ and\ \bibinfo {author} {\bibfnamefont {E.}~\bibnamefont
  {{Babaev}}},\ }\href {\doibase 10.1103/PhysRevLett.107.197001} {\bibfield
  {journal} {\bibinfo  {journal} {Phys. Rev. Lett.}\ }\textbf {\bibinfo
  {volume} {107}},\ \bibinfo {eid} {197001} (\bibinfo {year}
  {2011})}\BibitemShut {NoStop}%
\bibitem [{\citenamefont {{Garaud}}\ \emph {et~al.}(2013)\citenamefont
  {{Garaud}}, \citenamefont {{Carlstr{\"o}m}}, \citenamefont {{Babaev}},\ and\
  \citenamefont {{Speight}}}]{Garaud2013chiral}%
  \BibitemOpen
  \bibfield  {author} {\bibinfo {author} {\bibfnamefont {J.}~\bibnamefont
  {{Garaud}}}, \bibinfo {author} {\bibfnamefont {J.}~\bibnamefont
  {{Carlstr{\"o}m}}}, \bibinfo {author} {\bibfnamefont {E.}~\bibnamefont
  {{Babaev}}}, \ and\ \bibinfo {author} {\bibfnamefont {M.}~\bibnamefont
  {{Speight}}},\ }\href {\doibase 10.1103/PhysRevB.87.014507} {\bibfield
  {journal} {\bibinfo  {journal} {Phys. Rev. B}\ }\textbf {\bibinfo {volume}
  {87}},\ \bibinfo {eid} {014507} (\bibinfo {year} {2013})}\BibitemShut
  {NoStop}%
\bibitem [{\citenamefont {{Villain, J.}}(1975)}]{Villain1975}%
  \BibitemOpen
  \bibfield  {author} {\bibinfo {author} {\bibnamefont {{Villain, J.}}},\
  }\href {\doibase 10.1051/jphys:01975003606058100} {\bibfield  {journal}
  {\bibinfo  {journal} {J. Phys. France}\ }\textbf {\bibinfo {volume} {36}},\
  \bibinfo {pages} {581} (\bibinfo {year} {1975})}\BibitemShut {NoStop}%
\bibitem [{\citenamefont {Motrunich}\ and\ \citenamefont
  {Vishwanath}(2008)}]{Motrunich2008}%
  \BibitemOpen
  \bibfield  {author} {\bibinfo {author} {\bibfnamefont {O.~I.}\ \bibnamefont
  {Motrunich}}\ and\ \bibinfo {author} {\bibfnamefont {A.}~\bibnamefont
  {Vishwanath}},\ }\href@noop {} {\bibfield  {journal} {\bibinfo  {journal}
  {arXiv}\ } (\bibinfo {year} {2008})},\ \Eprint
  {http://arxiv.org/abs/0805.1494v1} {0805.1494v1} \BibitemShut {NoStop}%
\bibitem [{\citenamefont {Herland}\ \emph {et~al.}(2013)\citenamefont
  {Herland}, \citenamefont {Bojesen}, \citenamefont {Babaev},\ and\
  \citenamefont {Sudb\"o}}]{Herland2013}%
  \BibitemOpen
  \bibfield  {author} {\bibinfo {author} {\bibfnamefont {E.~V.}\ \bibnamefont
  {Herland}}, \bibinfo {author} {\bibfnamefont {T.~A.}\ \bibnamefont
  {Bojesen}}, \bibinfo {author} {\bibfnamefont {E.}~\bibnamefont {Babaev}}, \
  and\ \bibinfo {author} {\bibfnamefont {A.}~\bibnamefont {Sudb\"o}},\ }\href
  {\doibase 10.1103/PhysRevB.87.134503} {\bibfield  {journal} {\bibinfo
  {journal} {Phys. Rev. B}\ }\textbf {\bibinfo {volume} {87}},\ \bibinfo
  {pages} {134503} (\bibinfo {year} {2013})}\BibitemShut {NoStop}%
\bibitem [{\citenamefont {Carlstr\"om}\ and\ \citenamefont
  {Babaev}(2015)}]{Carlstrom2015}%
  \BibitemOpen
  \bibfield  {author} {\bibinfo {author} {\bibfnamefont {J.}~\bibnamefont
  {Carlstr\"om}}\ and\ \bibinfo {author} {\bibfnamefont {E.}~\bibnamefont
  {Babaev}},\ }\href {\doibase 10.1103/PhysRevB.91.140504} {\bibfield
  {journal} {\bibinfo  {journal} {Phys. Rev. B}\ }\textbf {\bibinfo {volume}
  {91}},\ \bibinfo {pages} {140504 (R)} (\bibinfo {year} {2015})}\BibitemShut
  {NoStop}%
\bibitem [{\citenamefont {Binder}(1981{\natexlab{a}})}]{Binder1981_critical}%
  \BibitemOpen
  \bibfield  {author} {\bibinfo {author} {\bibfnamefont {K.}~\bibnamefont
  {Binder}},\ }\href {\doibase 10.1103/PhysRevLett.47.693} {\bibfield
  {journal} {\bibinfo  {journal} {Phys. Rev. Lett.}\ }\textbf {\bibinfo
  {volume} {47}},\ \bibinfo {pages} {693} (\bibinfo {year}
  {1981}{\natexlab{a}})}\BibitemShut {NoStop}%
\bibitem [{\citenamefont {Binder}(1981{\natexlab{b}})}]{Binder1981_finitesize}%
  \BibitemOpen
  \bibfield  {author} {\bibinfo {author} {\bibfnamefont {K.}~\bibnamefont
  {Binder}},\ }\href {\doibase 10.1007/BF01293604} {\bibfield  {journal}
  {\bibinfo  {journal} {Z. Phys. B}\ }\textbf {\bibinfo {volume} {43}},\
  \bibinfo {pages} {119} (\bibinfo {year} {1981}{\natexlab{b}})}\BibitemShut
  {NoStop}%
\end{thebibliography}%

\end{document}